\documentclass[journal]{IEEEtran}
\usepackage{cite}

\ifCLASSINFOpdf
  \usepackage[pdftex]{graphicx}
  \DeclareGraphicsExtensions{.pdf,.jpeg,.png}
  \usepackage{epstopdf}
\else

\fi

\usepackage{amsmath}
\usepackage{amsfonts}
\usepackage{amssymb}
\usepackage{bm}
\usepackage{relsize}
\DeclareMathOperator{\Tr}{Tr}

\usepackage{hyperref}
\usepackage{cleveref}
\usepackage[dvipsnames]{xcolor}

\usepackage[normalem]{ulem}
\usepackage[defaultcolor=red]{changes}
\definechangesauthor[name={Qian}, color=red]{qs}

\usepackage{booktabs}

\usepackage{algorithmic}

\usepackage{array}

\ifCLASSOPTIONcompsoc
  \usepackage[caption=false,font=normalsize,labelfont=sf,textfont=sf]{subfig}
\else
  \usepackage[caption=false,font=footnotesize]{subfig}
\fi

\usepackage{epsfig}

\begin{document}
\title{Biomass Estimation and Uncertainty Quantification from Tree Height}

\author{Qian~Song,~\IEEEmembership{Member,~IEEE,}
Conrad~M~Albrecht,~\IEEEmembership{Member,~IEEE,}
Zhitong~Xiong,~
Xiao~Xiang~Zhu,~\IEEEmembership{Fellow,~IEEE}

\thanks{The work is jointly supported by the German Federal Ministry of Education and Research (BMBF) in the framework of the international future AI lab "AI4EO -- Artificial Intelligence for Earth Observation: Reasoning, Uncertainties, Ethics and Beyond" (grant number: 01DD20001) and by German Federal Ministry for Economic Affairs and Climate Action in the framework of the "national center of excellence ML4Earth" (grant number: 50EE2201C). \\ \emph{(Corresponding author: Xiao Xiang Zhu)}}
\thanks{Qian Song, Zhitong Xiong and Xiao Xiang Zhu are with the Chair of Data Science in Earth Observation (SiPEO), Technical University of Munich (TUM), Germany (e-mails: xiaoxiang.zhu@tum.de).

Conrad M Albrecht is with the Remote Sensing Technology Institute (IMF), German Aerospace Center (DLR), Germany.
}
\thanks{\copyright2023 IEEE. Personal use of this material is permitted. Permission from IEEE must be obtained for all other uses, in any current or future media, including reprinting/republishing this material for advertising or promotional purposes, creating new collective works, for resale or redistribution to servers or lists, or reuse of any copyrighted component of this work in other works. This material is referenced by \href{https://doi.org/10.1109/JSTARS.2023.3271186}{DOI:10.1109/JSTARS.2023.3271186}.
This is the pre-acceptance version, to read the final version please go to IEEE Journal of Selected Topics in Applied Earth Observations and Remote Sensing on IEEE XPlore.
}
}% 

\markboth{preprint of \href{https://doi.org/10.1109/JSTARS.2023.3271186}{DOI:10.1109/JSTARS.2023.3271186}}%
{Shell \MakeLowercase{\textit{Qian et al.}}: Biomass Estimation}

\maketitle
\begin{abstract}
We propose a tree-level biomass estimation model approximating allometric equations by LiDAR data.
Since tree crown diameters estimation is challenging from spaceborne LiDAR measurements,
we develop a model to correlate tree height with biomass on the individual tree level employing
a Gaussian process regressor. In order to validate the proposed model, a set of 8,342  samples on tree height, trunk diameter, and biomass has been assembled. It covers seven biomes
globally present. We reference our model to four other models based on both, the Jucker
data and our own dataset.
Although our approach deviates from standard biomass--height--diameter models, we demonstrate
the Gaussian process regression model as a viable alternative. In addition, we decompose
the uncertainty of tree biomass estimates into the model- and fitting-based contributions.
We verify the Gaussian process regressor has the capacity to reduce the fitting
uncertainty down to below 5\%. Exploiting airborne LiDAR measurements and a field
inventory survey on the ground, a stand-level (or plot-level) study confirms a low relative error of below 1\% for our model. The data used in this study are available at https://github.com/zhu-xlab/BiomassUQ.
\end{abstract}


\IEEEpeerreviewmaketitle

% -----------------------------------introduction-----------------------------------------------------
\section{Introduction}
\IEEEPARstart{A}{ccording} to the Food and Agriculture Organization (FAO), nearly 31\% of the global land surface
is covered by forests \cite{wall-to-wall}. Woodland is a valuable resource on Earth. Among others, it regulates the circulation
of air and water. According to EIP-AGRI\footnote{\url{https://ec.europa.eu/eip/agriculture/}}, 
forests in Europe provide around three million jobs, and forest biomass contributes to about half of the generated renewable energy \cite{eip2016}. Above-ground biomass (AGB) is a biophysical parameter for the total amount of accumulated organic material in an ecosystem. It has been applied widely as an index of forest volume. Moreover, it is key to
monitoring ecosystems and modeling climate change \cite{wall-to-wall, xiong2022earthnets}. In addition, biomass provides a tool to
evaluate carbon sequestration. Accurate biomass estimates help assess loss caused by wildfire \cite{singh2021analysis}.

\begin{figure*}[!t]
\centering
\includegraphics[width=6in]{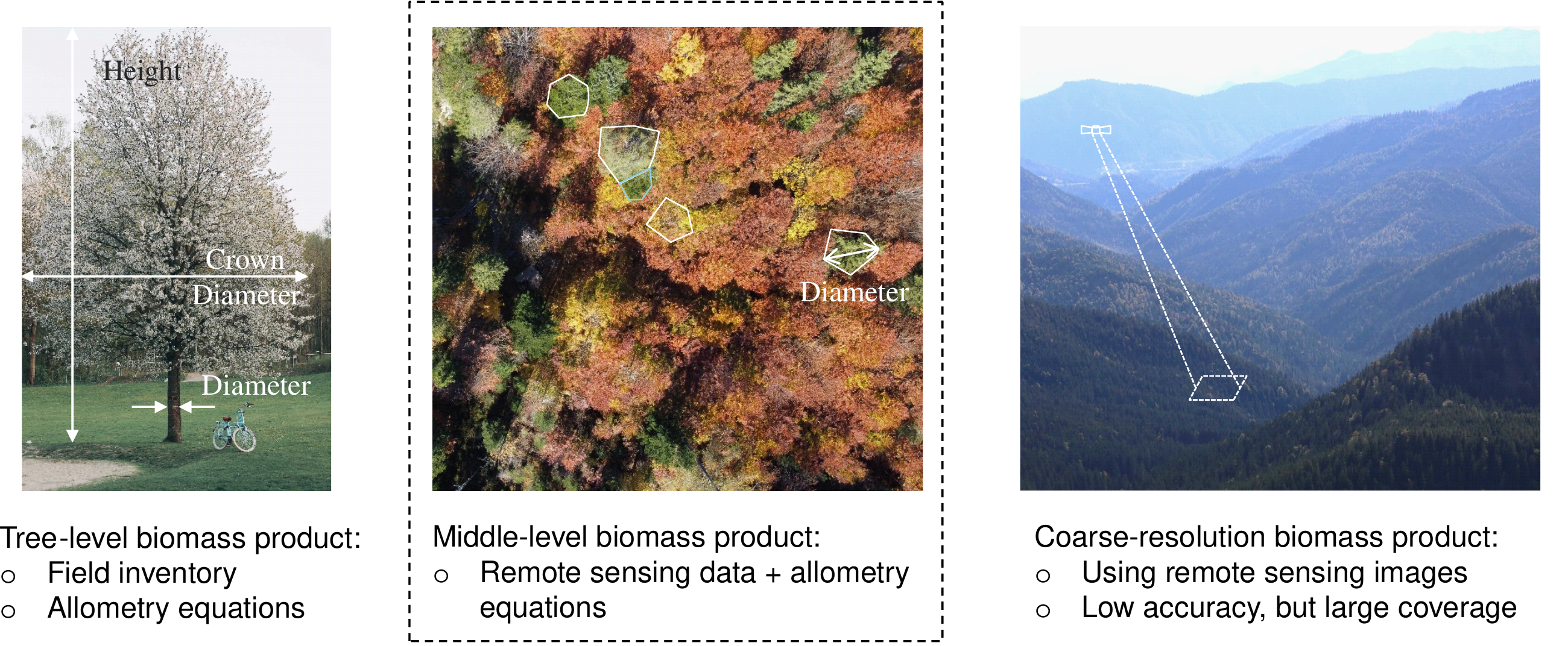}
\caption{Illustration of biomass products with fine (individual), middle, and coarse-grained (stand) levels.}
\label{motivation}
\end{figure*}

Forest AGB evaluation is categorized into three levels: fine, middle, and coarse-grained, cf.\ Fig.\ \ref{motivation}.
In situ measurements of biomass include tree harvest and desiccation to weigh the wood on a scale \cite{protocol}.
Despite errors in the harvesting process and scale uncertainties, this method most accurately evaluates biomass
on the tree level. However, such a destructive approach is costly in labor and time. As an alternative, allometric equations
provide estimates of biomass on individual tree level \cite{malhi2006regional, pan2011large, chave2014improved, anderson2015ctfs}.
Tree biomass is considered a function of tree height, trunk diameter at breast height (also known as trunk diameter
or simply diameter), wood density, crown diameter, etc. Luo et al.\ review state-of-the-art allometric
equations applied to tree species in China \cite{Allometry-review}. However, the method presented is not easily
transferred to a global scale. 

To generate large-scale (regional to global scale) biomass maps, spaceborne hyperspectral and synthetic aperture radar (SAR) sensors with wide swath and global or near-to-global coverage, have been applied in the literature \cite{carreno2020above, santoro2011retrieval, hayashi2019aboveground, schlund2019comparison}. Based on a regression model trained with ground reference and remote
sensing data pairs, the total amount of biomass for each pixel in the remote sensing imagery can get estimated.
However, limited by the coarse-grained resolution of remote sensing imagery\footnote{For example, 25 m spatial resolution
of spaceborne LiDAR sensor GEDI, \cite{GEDI, fusion, BIOMASS}}, ground truth reference data is costly to collect. According to
\cite{FVA} approximately $234$ trees per ha grow in the Black Forest, Germany.
Also, since the values of pixels in remote sensing
imagery are not related to trees' biomass one-to-one, the relative errors might raise up to 37\% \cite{wall-to-wall}.

\begin{figure*}[!t]
\centering
\subfloat[LiDAR return count statistics]{\includegraphics[width=3.3in]{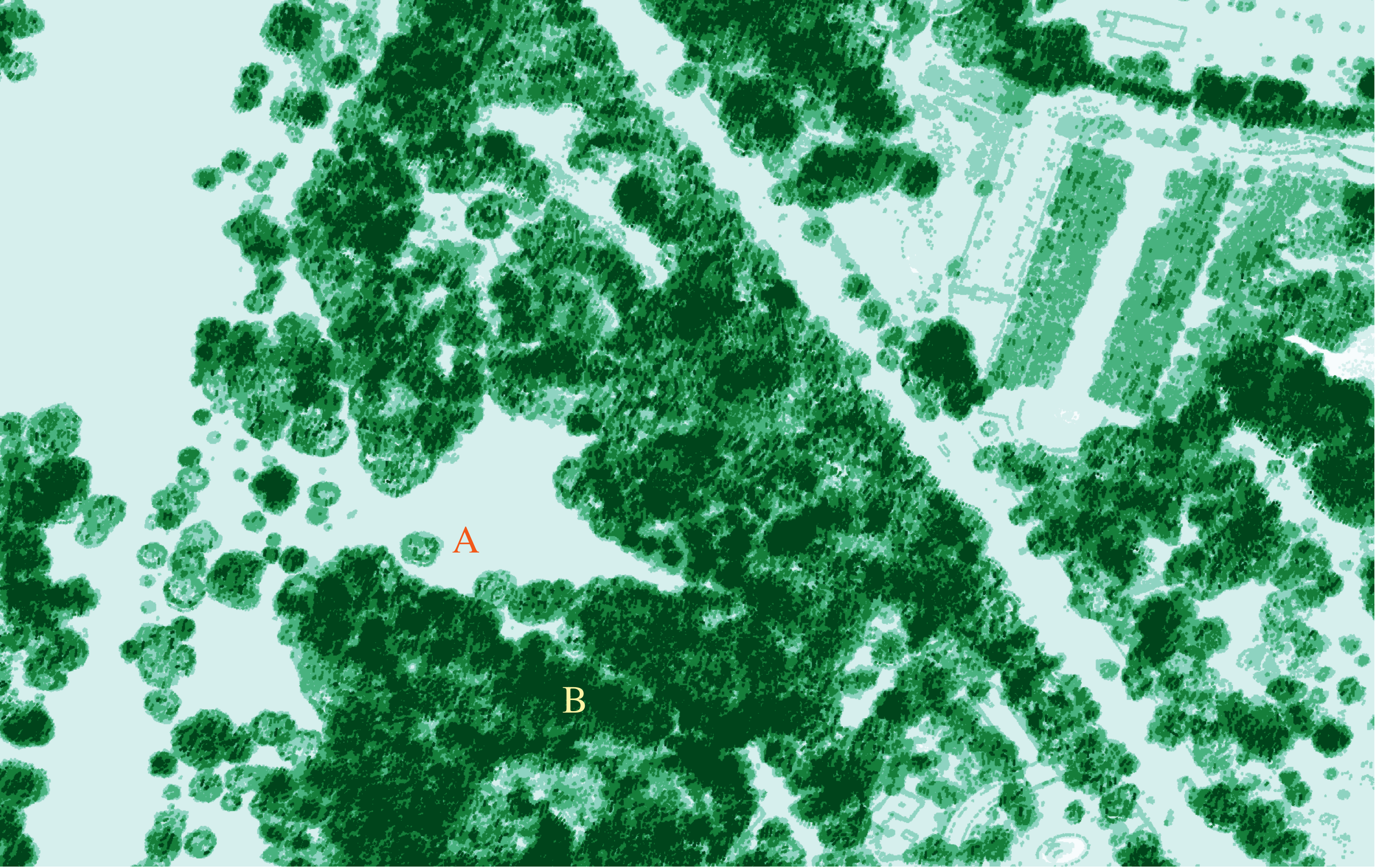}%
\label{fig1a}}
\hfil
\subfloat[RGB-color coded aerial imagery (NAIP, USDA Farm Service) with one-to-one correspondence of geospatial area in (a)]{\includegraphics[width=3.3in]{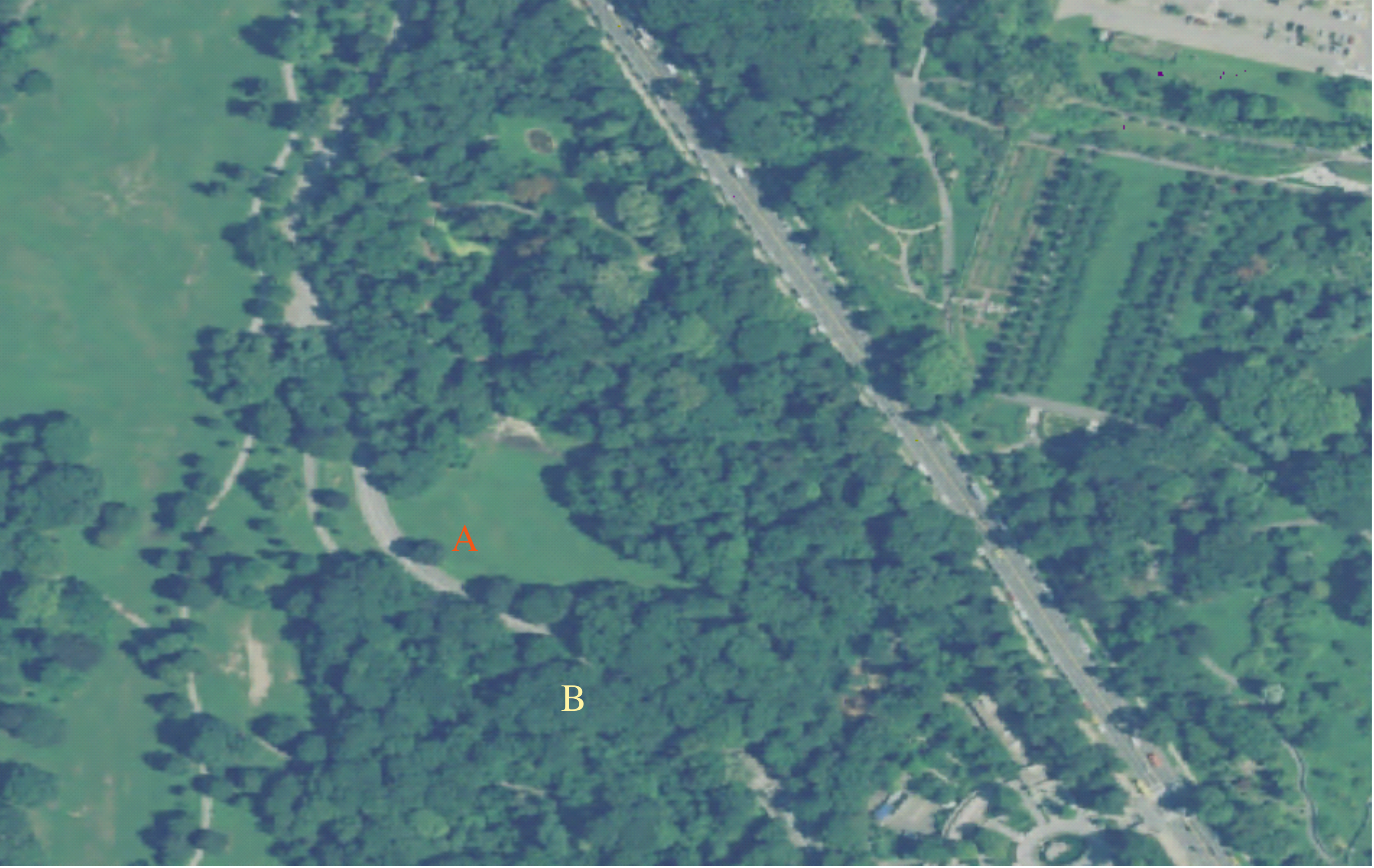}%
\label{fig1b}}
\caption{Sample of rasterized statistics of LiDAR return count (a) (for details in methodology cf.\ \cite{albrecht2021autogeolabel}). And
(b) depicts corresponding aerial imagery over Prospect Park in Brooklyn, New York, USA in 2017.
A and B mark areas of an isolated tree vs.\ a cluster of trees, respectively. LiDAR statistics and NAIP
imagery got harmonized by the Big GeoData platform PAIRS. \cite{klein2015pairs}}
\label{lidar}
\end{figure*}

A trade-off compared with in-situ biomass measurements is the estimation of tree-level parameters (such as height, crown diameter, etc.)
from high-resolution remote sensing data as input to allometric equations \cite{Yao-2012, Duncanson-2014, Jucker, klein2021quantification}.
Although highly correlated with biomass, parameters such as wood density and diameter cannot reliably get estimated
by aerial imagery. Jucker et al.\ confirmed that the height and crown diameter of trees are sufficient to estimate the
trunk diameter by a single equation. Crown diameter and height are easily derived from airborne laser scanning (ALS) data
\cite{Asner-2014a, Asner-2014b, Yao-2012, Duncanson-2014, Shendryk-2016}. However, the crown diameter estimation is a source of significant model error. Deciduous tree's crown changes with season, and extraction of the crown profile for individual
trees in dense forests is a major challenge. As demonstrated by \Cref{lidar}, the crown profile of isolated trees may
get reliably estimated, cf.\ label A. However, in densely populated areas such as labeled B, a reliable separation is close
to impossible. The analysis of high-resolution aerial imagery in \cite{H-CD-estimate} revealed the accuracy of estimated crown
diameter significantly varied with plots (0.63 and 0.85) when compared with height estimations. According to \cite{weiser2021tuaa}, the relative error of tree crown diameter derived from airborne/UAV-borne LiDAR data is significantly larger than that of derived tree height (19.22\% and 20.7\% for crown diameter, 11.70\% and 10.97\% for tree height estimation respectively). In fact, the crowns of
individual trees in e.g.\ rainforests may significantly intersect. 

 A few studies investigated stand-level height-biomass allometry \cite{mette2003height, caicoya2016large, choi2021improving}. Due to the lack of tree density information, these methods either focused on regional biomass estimation \cite{mette2003height}, or additional metrics such as percentile heights \cite{solberg2017biomass} and horizontal structure index \cite{choi2021improving} are integrated in the allometric relationships. Alternatively, tree-level height-biomass allometry may be beneficial as it directly includes tree density information. But as indicated in \cite{sullivan2018field}, height-diameter relationship varies even within a small scale given the compositional diversity. Therefore, to build a general tree-level height-biomass allometric relationship for fast biomass estimation, it is necessary: 1) the relationship is derived based on a dataset that collected over large areas; 2) the regression model should be less biased, so that aggregation of individual tree biomass within homogeneous forests would reduce biomass errors. In this paper, we evaluate such an approach:
We assemble a ground truth dataset including 8,342 measurements of tree height, trunk diameter, and biomass drawn from
global sampling. We proposed a Gaussian process regressor (GPR) which is a noise-aware model to capture the nonlinear relationship of biomass
and tree height and reduce the estimation bias. 

In addition, uncertainty quantification is crucial for evaluating the confidence level of derived models. Uncertainty related on plot-level biomass estimation based on tree allometry mainly comes from two sources \cite{protocol}: independent tree variables derivation uncertainty due to inaccurate and/or insufficient measurements and estimation uncertainty led by residual model noise and imperfect reference data. In this paper, we focus on the estimation uncertainty. Most existing publications reported their model's residual noise by metrics such as root mean square root (RMSE) \cite{wall-to-wall, Jucker}. On the other hand, Monte Carlo simulation approach was used to quantify the parameter uncertainty by selecting different samples as training data \cite{McRoberts2016}. Usually, they assume that the reference data are noise-free. However, due to the imperfect sampling strategy, the impact of reference data on the uncertainty quantification is non-negligible \cite{protocol}.
In this paper, we propose an algorithm to decompose the
estimation uncertainty into model uncertainty and fitting uncertainty. This way, we analyze the model uncertainties
of four allometric equations in contrast to the fitting uncertainties of five candidate models.
Stand-level uncertainties of the proposed GPR and two other models is evaluated based on LiDAR measurements and field
surveys.

% ---------------------------------------------dataset-----------------------------------------------------
\section{Datasets}
\label{datasets}
Our experiments employ a dataset collected from the open-source allometry databases \cite{BAAD, chave2014improved, 14}. Dataset provided by \cite{chave2014improved} consists of 4004 tree measurements at 58 sites in tropical forests over the globe. The Biomass And Allometry Database (BAAD) \cite{BAAD} includes 258,526 measurements over the globe collected from 175 studies. Each measurement records the tree height, tree components' biomass, etc. The biomass data collected in \cite{14} includes 6,604 records of trees in Eurasian forests (mainly in Russia). Then part of the measurements in the datasets were removed so that the left ones obey the following characteristics:
\begin{itemize}
    \item the total height and trunk diameter of the tree are recorded;
    \item the geographic location of the tree is recorded;
    \item the tree is harvested to measure its biomass;
    \item the tree's diameter exceeds 5 centimeters (cm);
    \item the tree's biomass passes the threshold of 2 kg.
\end{itemize}

Information on ecoregions defining geospatial boundaries of biome types (TEOW) has been downloaded from the world wide life fund (WWF) \cite{olson2001terrestrial}. According to geographic location, each measurement is
allocated by one of the seven biome types: tropical and subtropical forests, temperate mixed forests,
temperate coniferous forests, boreal forests, grasslands and shrublands, tundra, savannas,
woodlands, and mediterranean forests, or deserts and xeric shrublands.
Based on previous research, the parameters of the allometric equations depend on species,
climate, and environmental conditions. Instead of training multiple models for each species
and ecoregion, we explored the potential of using a single model to capture the variation caused by ecological factors.

\begin{figure*}[!t]
\centering
\subfloat[]{\includegraphics[width=7in]{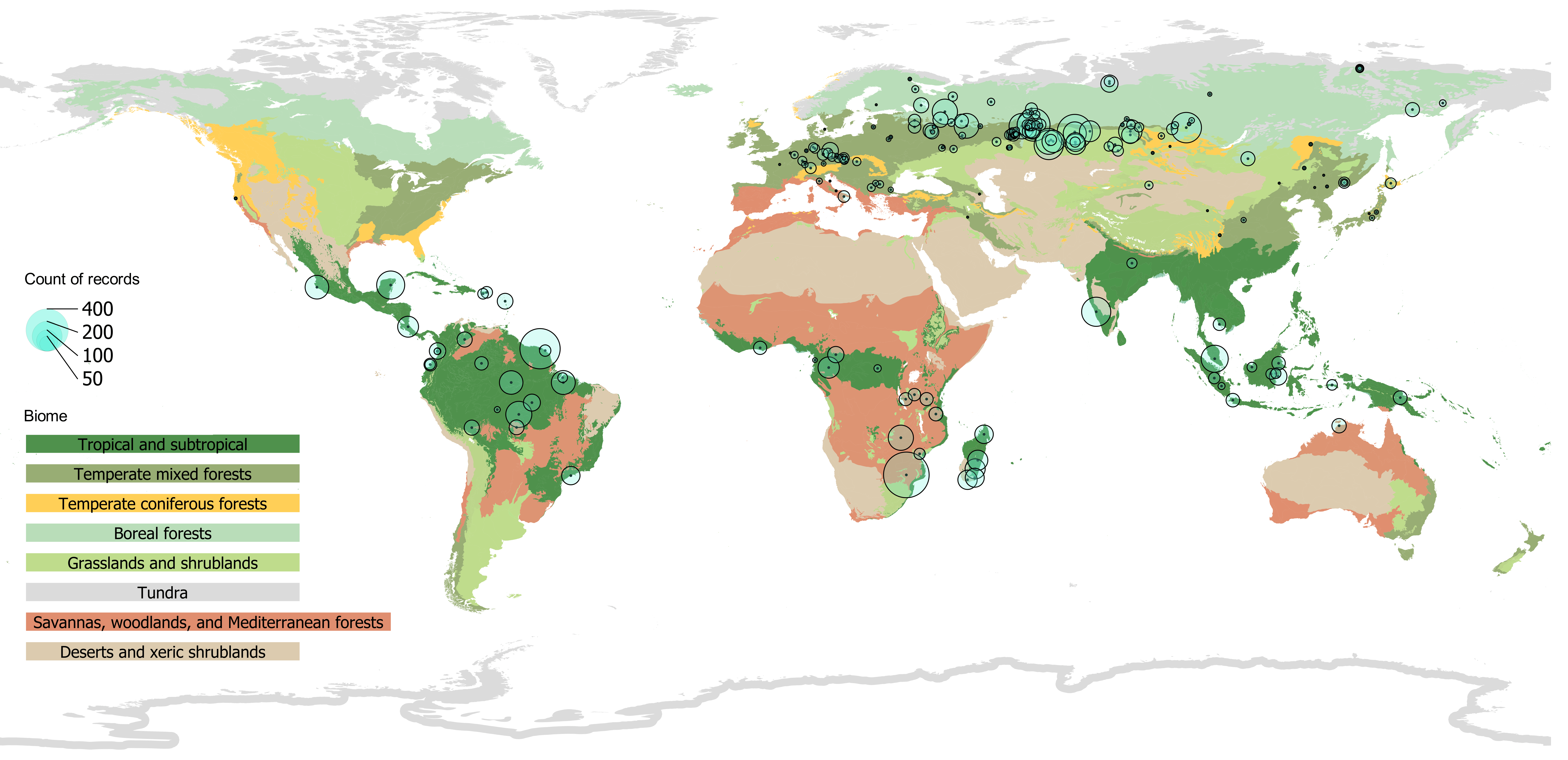}
\label{test}}
\hfill
\subfloat[]{\includegraphics[height=1.7in]{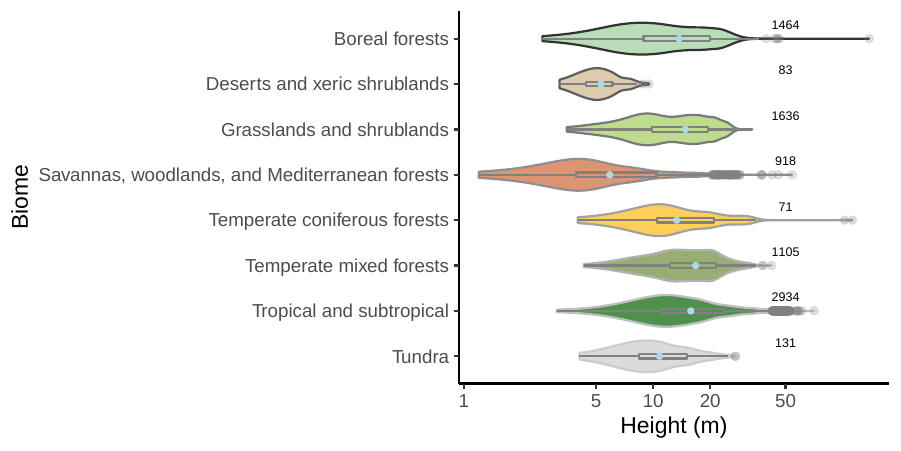}
\label{data_H}}
\subfloat[]{\includegraphics[height=1.7in]{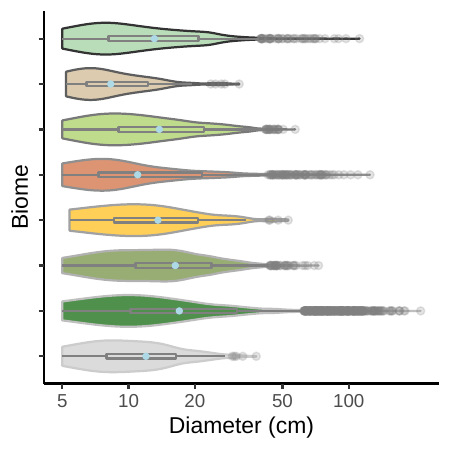}
\label{data_D}}
\subfloat[]{\includegraphics[height=1.7in]{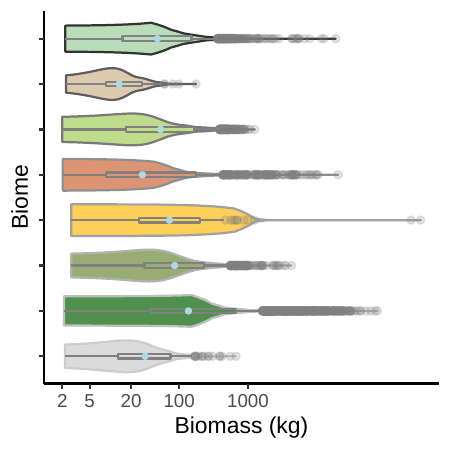}
\label{data_AGB}}
\caption{Summary of data collected \cite{BAAD, chave2014improved, 14}: Geospatial distribution of the measurements plotted on top of
the biome classification map. Circle diameters represent the number of records at each geo-location
(a); violin plots of the distributions of tree heights in meters (b), tree diameters in centimeters (c),
and above-ground biomass in kilograms (d) for various biomes. The number of records for each biome is
shown as a number to the right.}
\label{dataset}
\end{figure*}

\begin{figure}[!t]
\centering
\includegraphics[width=3.5in]{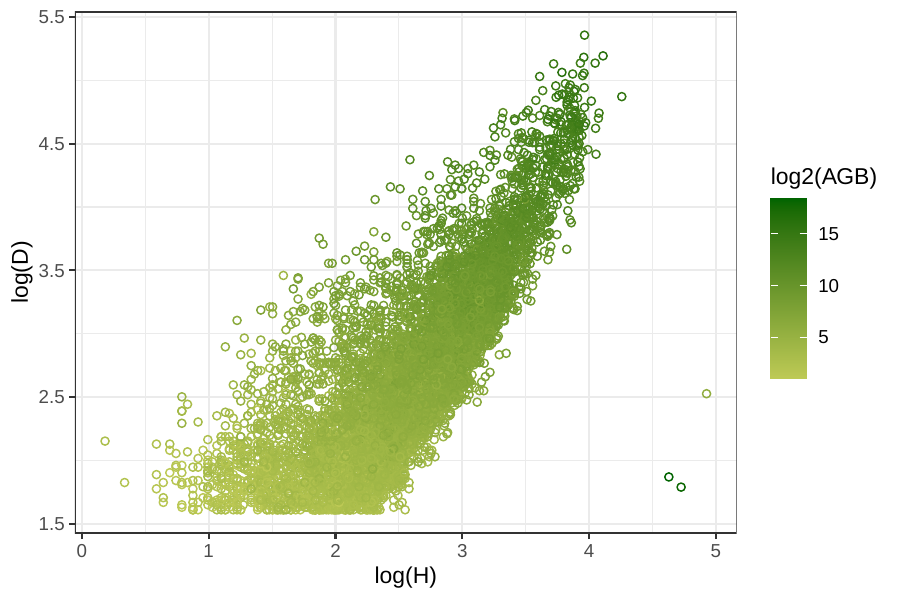}
\caption{Scatter plot of diameter and height in log-log scale, where the color stands for the AGB level.}
\label{AGB_H_D}
\end{figure}

Here we utilize a total number of 8,342 pairs of tree height-diameter-biomass measurements. In addition, we employ the Jucker data \cite{Jucker} as a reference for the proposed model. In addition to height, diameter, and biomass, the Jucker dataset \cite{Jucker} also records crown diameter. Although the Jucker data includes trunk diameter information, it contains 2,395 samples, only. \Cref{dataset} (a) visualizes the
geographic distribution and the number of records for various sites. We set the diameter of the blue circles in proportion to the number of measurements. Distinct biome regions are colored differently.
The dataset has global coverage containing all four forest types defined by the
\textit{Food and Agriculture Organization of the United Nations} (FAO) report\footnote{https://fra-data.fao.org/}.
\Cref{dataset} (b)--(d) present violin and box plots including median values (blue circles) and
outliers (grey circles) for each: height, diameter, and biomass of every biome type, respectively.
Figures in plot \Cref{dataset} (b) indicate the number of measurements in the biome regions. All three tree parameters
cover a wide range of measured values: biomass may get as little as two kilograms (kg), and it may
exceed 300 tons; tree height varies from 1.2 to 138 meters; tree diameters span a range from 5
cm up to more than two meters.
\Cref{AGB_H_D} plots the distribution of tree diameter $D$ vs.\ tree height $H$ in double-logarithmic
scale. Point colors indicate the log-scaled amount of biomass. Obviously, biomass increases
with tree height and trunk diameter. We observe: besides a small number of outliers, tree height and
trunk diameter are highly correlated.

%-------------------------------------method--------------------------------------
\section{Methodology}
\label{Methodology}
\subsection{Allometric Equation}
\label{allometry}

Biomass refers to the total amount of dry weight of organic material in a unit area, i.e.
the unit of biomass has dimension of e.g.\ kilograms per square meter $m^2$ or tons
(t) per hectare (ha). A tree's biomass accumulates from the biomass of stump, trunk,
branches, twigs, and leaves \cite{Stewart-1992}. To accurately measure the biomass of
trees, trees are felled, and dried at $105^{\circ}C$ for scaling. Large trees is impossible to
gauge. Instead, wood densities $\rho_i$ and volumes $V_i$ of all tree components labeled $i$
get recorded to estimate the biomass as $\sum_i \rho_iV_i$.
In \cite{Ketterings-2001, Basuki-2009, Maulana-2016}, the authors elaborated on the
process of dry biomass measurements.

The study \cite{Zhou-2009} demonstrates trunk biomass constitutes about 83\% of the
total biomass of the tree. In addition, based on measurements with total above-ground
biomass larger than 2 kg in BAAD \cite{BAAD}, leaf mass accounts for about 10\% of a tree's
biomass. Consequently, trunk biomass estimation needs the most attention. Assuming trunk biomass
modeled by a cone, the biomass $B=\rho V=\frac{1}{12}\pi\rho D^2H$, where $B$, $D$, and $H$
denote the tree's biomass, diameter, and height, respectively. $\rho$ is the average wood
density of dried trees that remote sensing is unable to capture. A central assumption of our
work reads: tree height is able to predict tree diameter such that biomass is predominantly
determined by tree height. Our experiments in \Cref{results} indicate complex relationships
beyond a log--log linear model. Hence, we establish a non-linear model $B\sim GP(H)$, where
$GP$ corresponds to a Gaussian process regressor detailed in \Cref{sec:GaussProcReg}.
The mapping of height $H$ to biomass $B$ is sensitive to average wood density and
the biome-dependent relationship of tree diameter vs.\ tree height.

\subsection{Gaussian Process Regressor}
\label{sec:GaussProcReg}

In order to model noisy biomass $B=B(H)$ depending on tree height measurements $H$
based on a set of samples $(H_1,B_1), (H_2,B_2),\dots,(H_n,B_n)$,
we employ Gaussian process regression.

Gaussian processes implement distributions over sequences of
variables $y'=(y'_1,y'_2,\dots,y'_n,y'_{n+1},\dots)$ fully parameterized by
mean values
\begin{equation}    
    \mu'_i=\langle y'_i\rangle,
\end{equation}
and the (symmetric) two-point correlation function
\begin{equation}
    K'_{ij}=\langle y'_iy'_j\rangle-\mu_i\mu_j=K'_{ji}\neq0,
\end{equation}
where we introduced the statistical averaging operator
\begin{equation}
\langle f\rangle=\int_{y'}f(y')\mathcal{N}(y'\vert\mu',K'),   
\end{equation}
over the Gaussian distribution $\mathcal{N}$.
Higher order (centralized) moments
\begin{equation}
    \langle y'_iy'_jy'_k\rangle,\quad\langle y'_iy'_jy'_ky'_l\rangle,\quad\dots
\end{equation}
can get expressed as products of two-point correlation functions \cite{isserlis1918formula}.
Thus, for the below
\begin{equation}
    y'\sim\mathcal{N}(\mu'=0,K'),
\end{equation}
samples sequences $y'$ from a multivariate Gaussian distribution
with zero mean and covariance matrix $K'$ defined by matrix elements
$K_{ij}$.

\begin{figure}[t]
\centering
\includegraphics[width=\columnwidth]{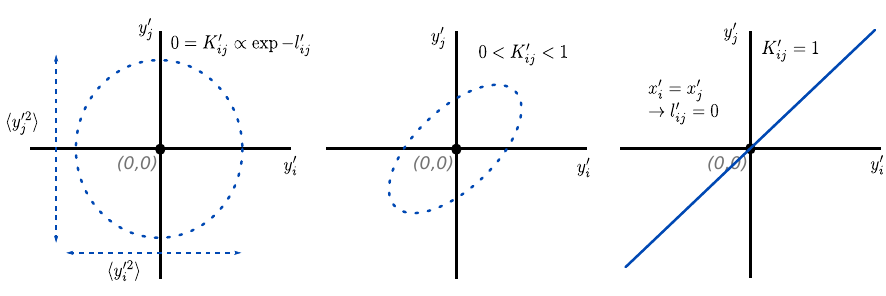}
\caption{\label{fig:CorrelatedGaussianNoise}
Illustration of correlated Gaussian noise and its relation to Gaussian process
regression models.}
\end{figure}

It is observed that in most physical systems (spatial) correlations exponentially decay
proportional to the length scale $l$, $\propto e^{-l}$. In fact, algebraic decays, i.e.\
$\propto l^{-\alpha}$, indicate strongly correlated systems close to phase transitions.
It is therefore reasonable to model the kernel
\begin{equation}
    K'_{ij}\propto\exp-(x'_i-x'_j)^2,
\end{equation}
where the $x'_i$ and $x'_j$ is associated with either height measurements $H$ or model
inputs $\hat{H}$. \Cref{fig:CorrelatedGaussianNoise} depicts the two-point correlation
for two measurements $y'_i=B_i$ and $y'_j=B_j$ far apart ($x'_i=H_i\ll H_j=x'_j$),
close ($H_i\approx H_j$), and identical ($H_i=H_j$).

In order to predict $N-n$ values $\hat{y}$ given a set of $n$ observation pairs
$(x_1,y_1),(x_2,y_2),\dots,(x_n,y_n)$ we cast both into an $N$-variate Gaussian distribution over
variables $y'=(y, \hat{y})$ and corresponding (given) parameters $x'=(x,\hat{x})$ to sample from
the resulting conditional Gaussian probability distribution\footnote{
    The derivation follows standard statistics textbooks such as in \cite{bishop2006pattern},
    section 2.3.1, where $\hat{K}=\Sigma_{aa}$, $K=\Sigma_{bb}$, and $\Sigma_{ab}=\kappa=\Sigma_{ba}^T$
    with $\mu_a=\mu_b=0$.
}
\begin{equation}
    \hat{y}\vert y\sim\mathcal{N}(\kappa K^{-1}y, \hat{K}-\kappa K^{-1}\kappa^T),
\end{equation}
where we decomposed the covariance matrix $K'$ according to $y'=(y,\hat{y})$:
\begin{equation}
K'= \left(
    \begin{array}{cc}
         K&\kappa  \\
         \kappa^T&\hat{K}
    \end{array}
\right),
\end{equation}
with $\kappa^T$ the transposed matrix of $\kappa$. Note that $K$ depends on $x$, only.
Similarly, $\hat{K}$ takes prediction inputs $\hat{x}$, only. In contrast, $\kappa$
entails a mix of $x$ and $\hat{x}$.

By design $K_{ii}=0$ such that we may flexibly add uncorrelated Gaussian measurement noise
to $x$ through
\begin{equation}
    K_{ij}\to K_{ij}+\sigma^2\delta_{ij},
\end{equation} where $\sigma\in\mathbb{R}$ quantifies
the variance of the uncorrelated measurement noise. The Kronecker-delta $\delta_{ij}$ turns
zero for all indices except for $i=j$ where it assumes the value $1$. In addition, we
may want to explicitly model a mean\slash\textit{expected} (biomass) function $m(x')=b(H')$
that translates into
\begin{equation}
    \kappa K^{-1}y \to m(\hat{x})+\kappa K^{-1}[y-m(x)],
\end{equation}
In summary: Given 
\begin{itemize}
    \item the one-dimensional radial basis function kernel
          \begin{equation}
              K'_{ij}=\exp-(H'_i-H'_j)^2/2l^2,
          \end{equation}
          at length scale $l$
    \item measured height values $H=(H_1,H_2,\dots,H_n)$ and model
          input values $\hat{H}$
    \item biomass data $B=(B_1,B_2,\dots,B_n)$
\end{itemize}
we statistically model $\hat{B}=B(\hat{H}\vert B,H)$ through a (mean-shifted) multi-variate
Gaussian distributed $\hat{B}\sim\mathcal{N}(\mu, \Sigma)$
\begin{enumerate}
    \item at \textbf{mean value} $\mu$
         \begin{eqnarray}
         \label{mean_predict}
            \mu-b_0          &=& \kappa (K+1\sigma)^{-1}(B-b_0),\\
            \log K_{ij}      &=&-(H_i-H_j)^2/2l^2,\\
            \log \kappa_{ij} &=&-(\hat{H}_i-H_j)^2/2l^2,
         \end{eqnarray}
         where $1$ denotes the unit matrix, $b_0$ is a constant hyper-parameter mean
         biomass, i.e.,
         \begin{equation}
             m(x')=b(H')=b_0=\text{const.},
         \end{equation}
         getting optimized alongside with $l$, and
    \item associated \textbf{covariance matrix} $\Sigma$   
         \begin{eqnarray}
         \label{cov_mat}
             \Sigma-\hat{K}   &=&-\kappa (K+1\sigma^2)^{-1}\kappa^T,\\
             \log\hat{K}_{ij} &=&-(\hat{H}_i-\hat{H}_j)^2/2l^2.
         \end{eqnarray}
\end{enumerate}
Note that the various elements of $K'$ contain both, data tree heights $H_i$ of
sample biomass terms $B_i$, and values $\hat{H}_i$ for biomass values $\hat{B}_i$ to predict.
Also, the constant offset $b_0$ could get replaced by a more generic (known) functional
dependence, e.g.\ a linear model $b(H')=b_1H'+b_0$, etc.

The scalar hyperparameters $l$ and $b_0$ get optimized by maximization of the
predictor variables $\hat{B}$-marginalized likelihood
\begin{eqnarray}
    B\vert\hat{B}\sim\hspace{-.2ex}p(B\vert H)&\hspace{-1ex}=\hspace{-1ex}&\int_{\hat{B}}p(B,\hat{B}\vert H,\hat{H})\nonumber\\
                  &\hspace{-1ex}\propto\hspace{-1ex}&{\frac{\exp-(B-b_0)^T(K+1\sigma^2)^{-1}(B-b_0)/2}{\sqrt{\det (K+1\sigma^2)}}}\nonumber,\\
\end{eqnarray}
i.e.\ when taking the logarithm, it is minimized the scalar (loss) function
\begin{eqnarray}
L(l,b_0) &=&\log p(B\vert H)\nonumber\\
%&\propto\sum_{i}\log(K_{ii}+\sigma^2)+\sum_{ij}(B_i-b_0)(K+1\sigma^2)^{-1}_{ij}(B_j-b_0)\nonumber\\
&\propto&\sum_i\beta^2_i/(k_i+\sigma^2)+\log(k_i+\sigma^2),
\end{eqnarray}
where we exploited $\log\det=\Tr\log$ \cite{bellman1997introduction}, and defined the
$l$-dependent eigenvalues $k_i$ of the symmetric matrix $K+1\sigma$. $\beta_i$ denotes
the $b_0$-dependent components of vector $B-b_0$ in the eigenbasis of $K$.

When using the Gaussian process for tree biomass estimation, the mean biomass offset can be predicted as a linear combination of observed biomass offsets weighted by the covariance matrix (closer samples have higher weights), as formulated in Eq. \ref{mean_predict}. In Eq. \ref{cov_mat}, the prediction uncertainty (covariance matrix) consists of two parts, the inherent noise level, and the height distances between the observed data and the prediction inputs.

Notice: Gaussian process regressors can get casted into the framework of non-parametric
Bayesian models. An approach that has proven efficient in many non-linear regression
tasks \cite{williams1996gaussian,camps2018physics,xiong2022doubly}.

\subsection{Evaluation Methods}
We compare the Gaussian process biomass-height model with a random forest (RF) model and three allometric equations,
specifically: biomass--height--crown diameter (LR), biomass--height (LR2), and biomass--height--diameter (LR3). Random forest is a data-driven non-linear regressor, which has been widely applied to biomass estimation \cite{wall-to-wall}. 
The form of the three allometric equations read:
\begin{eqnarray}
\label{BHCD}
\text{LR:} &\ln B=&a\ln (H \times CD) + b + \epsilon,\\
\label{BH}
\text{LR2:}&\ln B=&a\ln H + b + \epsilon,\\
\label{BHD}
\text{LR3:}&\ln B=&a\ln H + b\ln D + c+ \epsilon,
\end{eqnarray}
where $a$, $b$, $c$ are the coefficients and bias terms determined by the training data;
$CD$ refers to the crown diameter; and $\epsilon$ is model residuals. Since no crown diameter measurements in curated data in \cref{datasets}, we utilize an alternative biomass-diameter model:
\begin{equation}
\label{B-D}
\text{LR: }\ln B = a\ln (D) + b + \epsilon.
\end{equation}
\subsubsection{Tree-level Results Evaluation}
To evaluate model accuracy, three indices get derived: R--squared ($R^2$), root mean square error (RMSE),
and model bias. R--square refers to the coefficient of determination, and is defined according to
\begin{equation}
R^2(y, \hat{y}) = 1-\frac{\sum_{i=1}^{n}{(y_i-\hat{y}_i)^2}}{\sum_{i=1}^{n}{(y_i-\bar{y})^2}} = \frac{ESS}{TSS} = 1 - \frac{RSS}{TSS},
\end{equation}
where $y_i$ and $\hat{y}_i$ are the $i$-th ground truth and predicted values.
$\bar{y}$ amounts for the average mean of ground truth. ESS, TSS, and RSS abbreviate the
definitions of \textit{explained sum of squares}, \textit{total sum of squares}, and
\textit{residual sum of squares} in line with
\begin{eqnarray}
ESS &=& \sum_{i=1}^{n}{(\hat{y}_i-\bar{y})^2},\\
TSS &=& \sum_{i=1}^{n}{(y_i-\bar{y})^2},\\
RSS &=& \sum_{i=1}^{n}{(y_i-\hat{y}_i)^2}.
\end{eqnarray}

According to these definitions, the R--squared score may receive impact by a single, strongly
biased estimation. Thus, calculating $R^2$, we exclude outliers when the corresponding
absolute error exceeds the mean absolute error by at least three times, cf.\ red circles in \Cref{plotJucker}. RMSE is calculated as follows
\begin{equation}
RMSE(y, \hat{y}) = \sqrt{\frac{1}{n}\sum_{i=1}^{n}{(y_i-\hat{y}_i)^2}}.
\end{equation}
Bias relates to relative systematic error. It is defined as
\begin{equation}
Bias(y, \hat{y}) = \frac{1}{n}\sum_{i=1}^{n}{\frac{\hat{y}_i -y_i}{y_i}}.
\end{equation}
The negative or positive value of the bias indicates biomass under- or overestimation. 

In the following, we use a binning method to visualize prediction errors for input dimensions.
That is, the residuals between observed and predicted biomass $y_i - \hat{y}_i$ are calculated.
According to the percentile input values such as height get assigned on a logarithmic scale,
and residuals are split into separate groups (bins). The interval spanning the mean plus-minus half a
standard deviation for each bin is presented alongside the fitted curve.
The result visually depicts the mean and standard deviation of the prediction errors. In addition, it illustrates the level of over- or underestimation. 
\subsubsection{Plot-level Results Evaluation}
 We quantify uncertainty
on stand level by relative error (RE) and relative root mean square error (relative RMSE,
denoted as $\%RMSE$). Since in-situ data for biomass is unavailable, the biomass obtained by the LR3 model trained on our curated dataset using the filed inventoried tree heights and diameters serves as ground truth. 

By aggregating individual tree information, the relative error
denotes the ratio of the sum over residuals and the sum over predicted biomass values
by LR3: 
\begin{equation}
RE = \frac{\sum_{i=1}^{n} [LR3(H_i, D_i) - f(x_i)]}{\sum_i LR3(H_i, D_i)}.
\end{equation}
Here $H_i, D_i$ represent tree height and diameter of the $i$-th tree in the plot; $f$
indicates one of the candidate models, and $x_i$ signals model input parameter(s).

The \textit{relative RMSE} refers to the ratio of RMSE and the mean of biomass predicted by LR3:
\begin{equation}
\% RMSE = \frac{\sqrt{\frac{1}{N}\sum_{i=1}^{N}{(LR3_i - f_i)^2}}}{\sum_{i=1}^{N}{LR3_i}},
\end{equation}
where $i$ indexes the $i$-th plot.
\subsubsection{Uncertainty Evaluation}
\begin{figure}[t]
\centering
\includegraphics[width=3in]{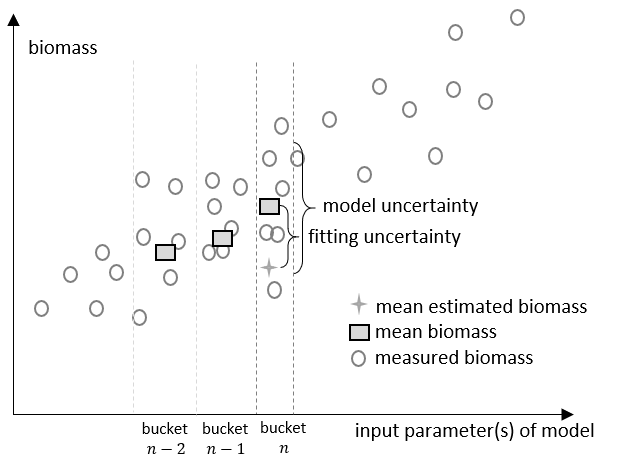}
\caption{Illustration of model uncertainty vs.\ fitting uncertainty. Dots and the solid lines
refer to the sample measurements and the averaged biomass, respectively. Two sources of
uncertainties we focus on: the model uncertainty that corresponds to the standard deviation
of the sample measurements, and the fitting uncertainty---that is: the deviation of averaged
biomass and the regressor-predicted biomass.}
\label{uncertainty}
\end{figure}

In the following, we elaborate on the uncertainty evaluation algorithm in use.
We concern with two sources of uncertainties, namely: model uncertainty and fitting uncertainty,
cf.\ \Cref{uncertainty}. Model uncertainty indicates variance rooted in model selection such
as the choice of input parameters for allometric equations, etc. In practice, tree biomass depends
on many factors such as annual rainfall, species, and average annual temperature. The model
at hand might bear the limited capacity to capture such dependencies. As a result, the mapping from input to biomass remains noisy with the model unable to capture
such residuals. We define model uncertainty by the variability of measured biomass.
Concerning the wide range and heavy tail of the tree biomass distribution, we work with
$\log$--scaled quantities. Specifically, the model uncertainty index is calculated
as follows: 1) measurements get sorted by the input parameter such as tree height,
crown diameter, etc., and are grouped into $n$ buckets; 2) for each of the $n$
groups, the ratio of standard deviation to mean of the biomass is calculated on a logarithmic scale; 3) the overall model uncertainty is calculated as the averaged ratios.
In some allometric models, biomass is correlated with multiple parameters such as height and
diameter. Subsequently, the measurements are sorted by one of the parameters.

In general, increasing the number of input parameters has the potential to decrease model
uncertainty at the price of additional effort to collect data. An alternative
provides training separate models for each biome, species, and age. On the downside,
this approach requires vast amounts of in-situ measurements harvesting trees.
Also, there exists an option to reduce model uncertainty for stand-level products
where spatial aggregation of tree biomass may cancel over- and underestimation \cite{wall-to-wall}.

When employing various forms of regression models,
the fitting precision of the regressors varies. We refer to the discrepancy in average
biomass and predicted biomass as fitting uncertainty. Computation of the fitting uncertainty
is similar to model uncertainty calculation: 1) the measurements are sorted by input
parameter and assigned into $n$ evenly spaced pockets; 2) the absolute error of
predicted biomass subtracted by the mean observed biomass is determined for each
measurement of every group; 3) for each of the $n$ groups, the ratio of mean absolute
error (MAE) to mean observed biomass is computed on a logarithmic scale; 4) the overall fitting uncertainty is computed as an averaged ratio.

We are going to demonstrate that Gaussian process regressors reduce the fitting uncertainty with regard to the random forest and linear regression models.

%--------------------------------------------results------------------------------------------
\section{Experiments and results}
\label{results}
\subsection{Jucker Data}

We adopt the Jucker data in order to benchmark the Gaussian process regressor in reference
to the other models introduced. The dataset includes 2,395 measurements including records
on crown diameter. In order to compare with and validate trained models such as LR proposed
by Jucker et al., we filter training data to exclude diameters smaller than $5$ cm.
We apply a random split into training and test sets in proportion to 9:1. 

Table \ref{indexJucker} summarizes the performance of the five regressors. LR3, the biomass-height-diameter
in \Cref{BHD} performs best in terms of all three indicators picked. $R^2$, $RMSE$, and $Bias$
yield values 0.95, 424.68 kg, and 0.08, respectively. As detailed in \Cref{allometry}, a linear
model on a log scale is insufficient to fit tree height--biomass data, and the LR2 model
performance is reflected by an increased RMSE (1.47 Mg), the most prominent bias (0.29),
and an $R^2$ score equating to 0.53. Non-linear models---such as random forest and
Gaussian process regressor---reduce the RMSE to 1.15 Mg and 1.12 Mg, respectively.

Our experiment indicates a low RMSE for the LR, RF, and GPR models, namely: 1.11 Mg,
1.15 Mg, and 1.12 Mg. Compared with LR \ref{BHCD} (the most widely used model), RF and GPR yield higher R--square scores by margins
of 21\% and 27\%, respectively. Consistently, the bias drops by 19\% and 15\%. Based on the above findings, our tree height--only Gaussian process regressor provides a
serious option for biomass modeling when compared with state-of-the-art biomass-height-crown
diameter models. 

\begin{table}[]
\footnotesize
\renewcommand{\arraystretch}{1.35}
\caption{Summary of R-square scores, RMSE and Bias of a series of regression models for
biomass estimation benchmarked on the Jucker data}
\label{indexJucker}
\centering
\setlength{\tabcolsep}{16pt}
\begin{tabular}{cccc}
\toprule
             & \textbf{R2} & \textbf{RMSE (kg)} & \textbf{Bias} \\ \hline \\[-1.2em]
\textbf{LR}  & 0.664463    & 1108.855      & 0.26314       \\
\textbf{LR2} & 0.53256     & 1466.553      & 0.2931        \\
\textbf{LR3} & 0.950643    & 424.6752      & 0.079669      \\
\textbf{RF}  & 0.803928    & 1147.066      & 0.20911       \\
\textbf{GPR} & 0.837668    & 1117.9        & 0.218732      \\
\bottomrule
\end{tabular}
\end{table}

\begin{figure*}[!t]
\centering
\subfloat[]{\includegraphics[height=1.5in]{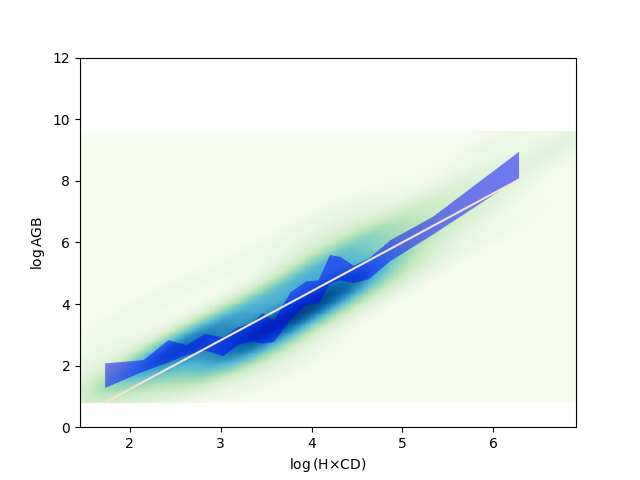}
}
\subfloat[]{\includegraphics[height=1.5in]{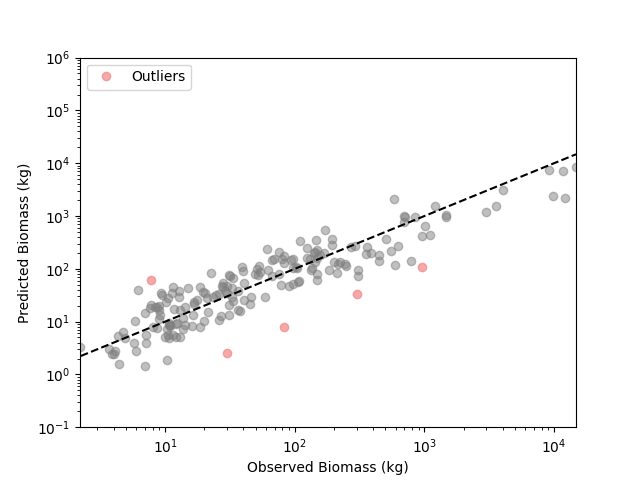}
}
\subfloat[]{\includegraphics[height=1.5in]{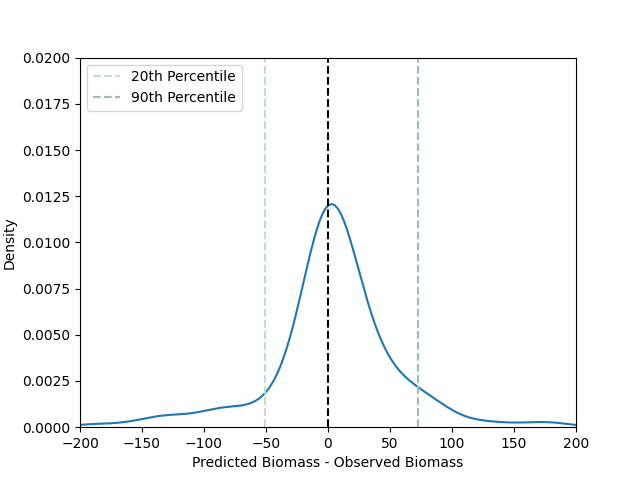}
}
\hfil
\subfloat[]{\includegraphics[height=1.5in]{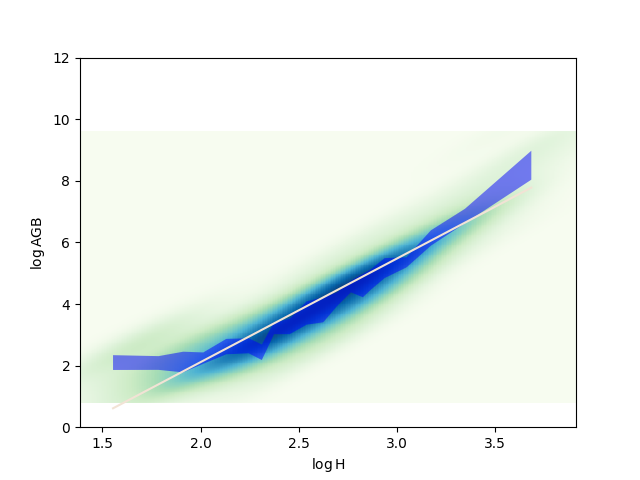}
}
\subfloat[]{\includegraphics[height=1.5in]{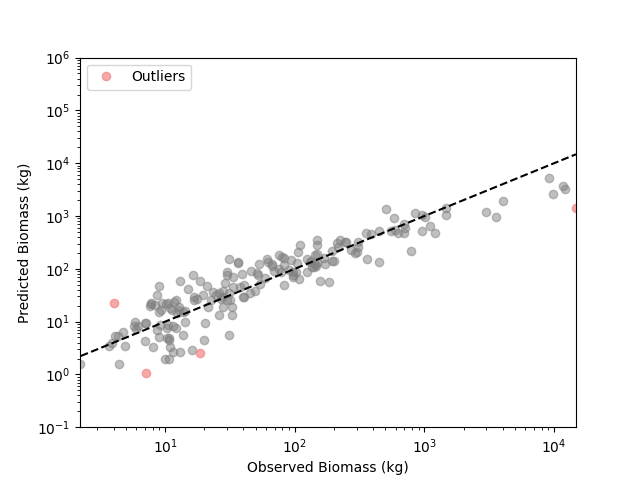}
}
\subfloat[]{\includegraphics[height=1.5in]{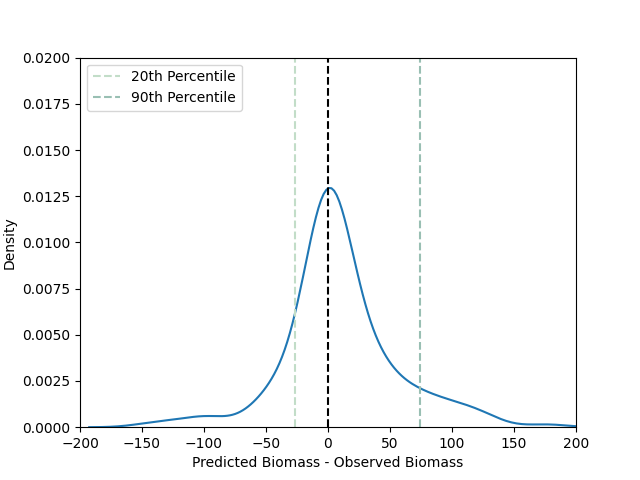}
}
\hfil
\subfloat[]{\includegraphics[height=1.5in]{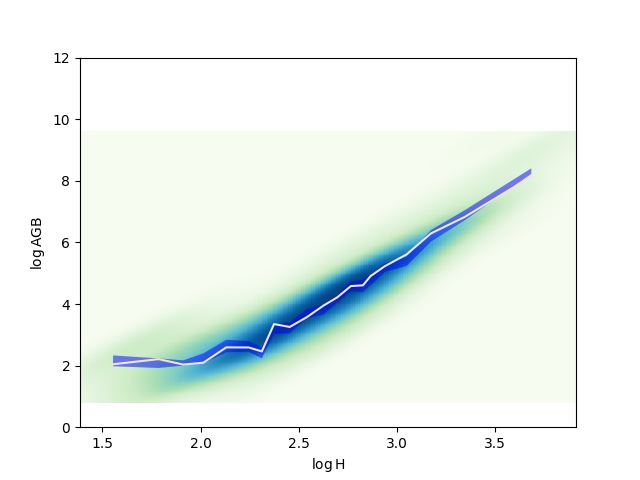}
}
\subfloat[]{\includegraphics[height=1.5in]{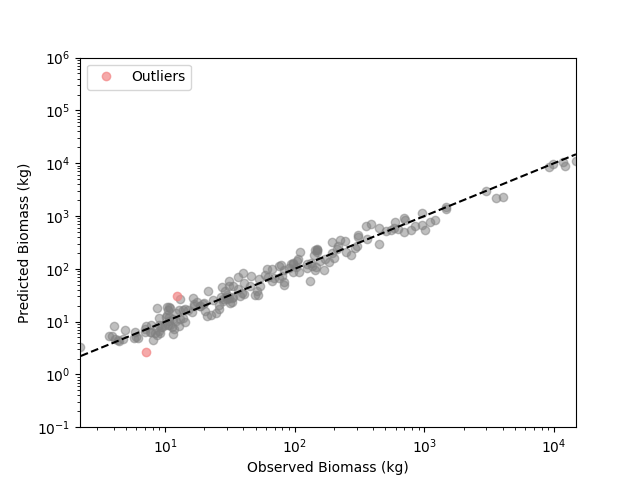}
}
\subfloat[]{\includegraphics[height=1.5in]{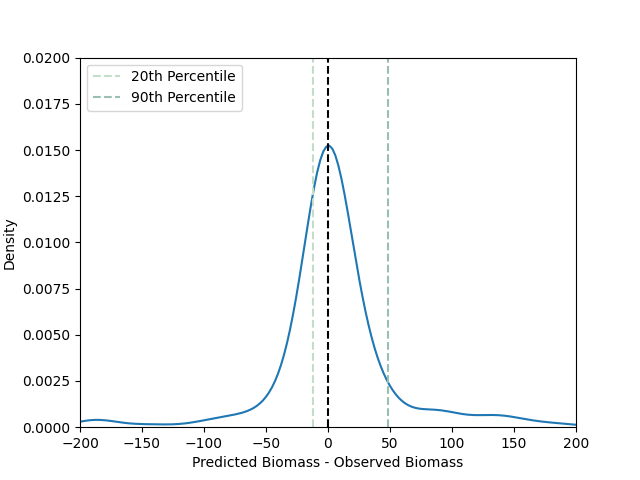}
}
\hfil
\subfloat[]{\includegraphics[height=1.5in]{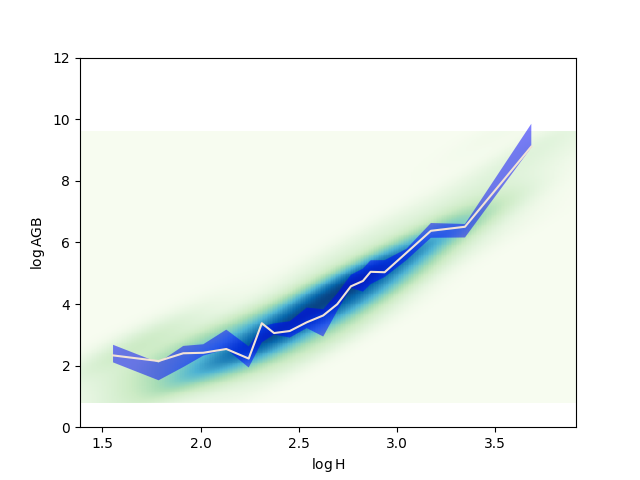}
}
\subfloat[]{\includegraphics[height=1.5in]{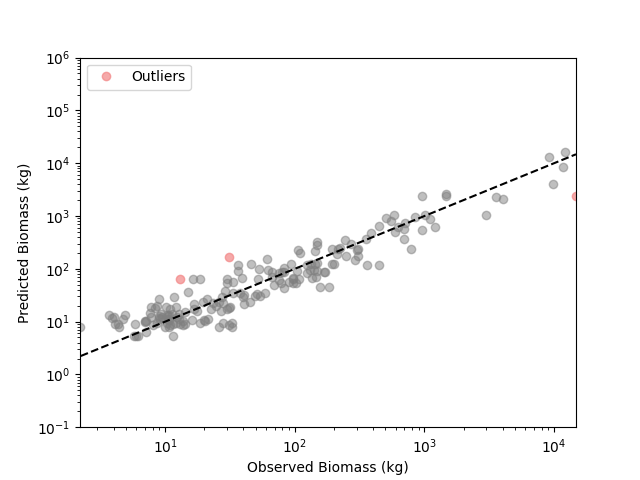}
}
\subfloat[]{\includegraphics[height=1.5in]{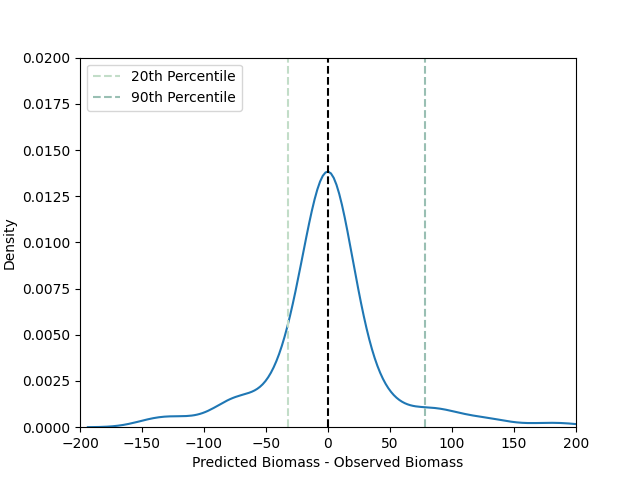}
}
\hfil
\subfloat[]{\includegraphics[height=1.5in]{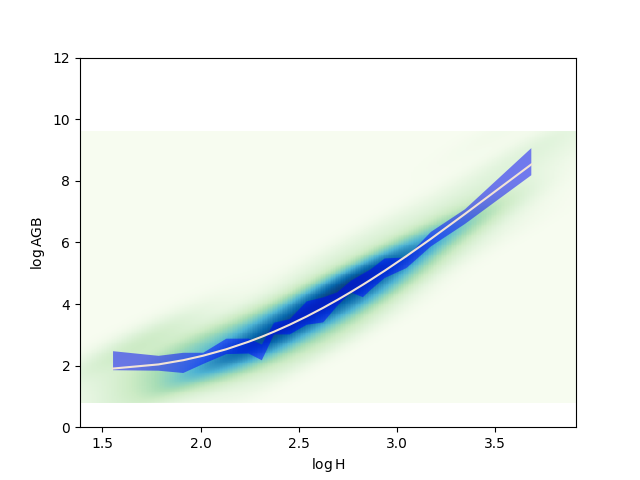}
}
\subfloat[]{\includegraphics[height=1.5in]{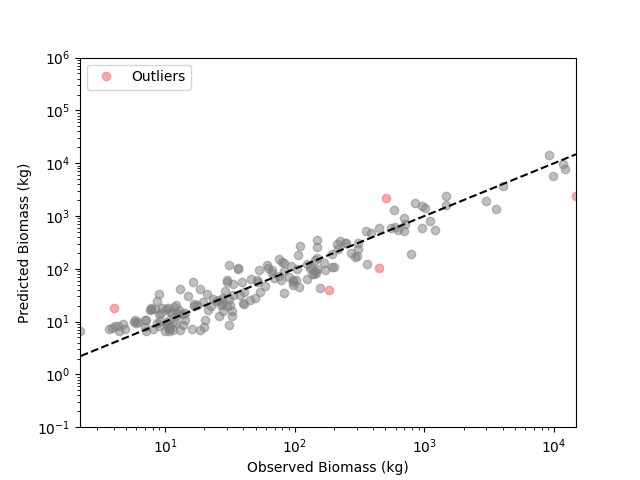}
}
\subfloat[]{\includegraphics[height=1.5in]{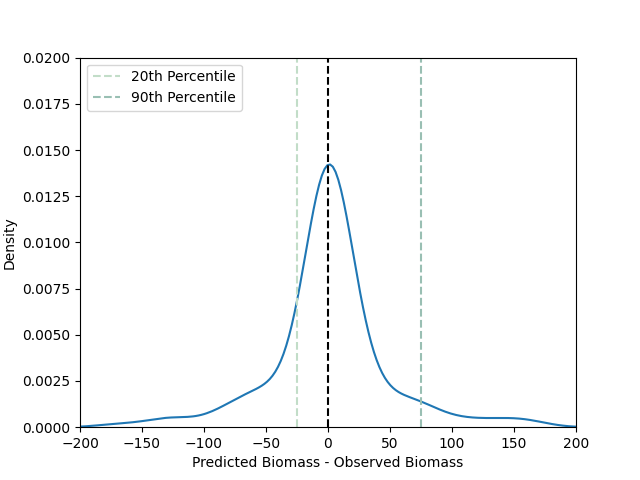}
}
\caption{\label{plotJucker} Plots of the fitted curves with corresponding prediction
errors (left column), scatters of predicted and observed biomass shown in middle column,
and the distributions of errors depicted by the contents of the right column.
The evaluation is based on the Jucker data. Each row corresponds to one of the five
models---from top to down: LR (a)-(c), LR2 (d)-(f), LR3 (g)-(i), RF (j)-(l), GPR (m)-(o).}

\end{figure*}

The left column in \Cref{plotJucker} lists fitted curves (blue lines) and corresponding errors distributions (blue areas) for the five models we did investigate. The background
resembles density maps of biomass-input parameter pairs. We observe the LR3 model fits
best with the data, it yields the lowest uncertainty, cf.\ it exhibits the most narrow
range of green-dashed, vertical lines in the plots of the right most column of \Cref{plotJucker}.
In \Cref{plotJucker} (a), although the fitted line doesn't align perfectly with the data, the actual biomass is linearly correlated with the product of tree height and crown diameter, which implies that a linear log-log model can describe the relationship between them.

In terms of single-parameter models, LR2 overestimates biomass predictions for medium range
tree height values, and it strongly underestimates the biomass for small and large heights---a
linear model does not properly capture the non-linear biomass--height relationship.
Both, the random forest and Gaussian process regressor render well with the data. However,
the fitted curve of the RF model is less regular compared with GPR bearing risk of less
robustness with respect to outliers.

The scatter plots of \Cref{plotJucker} (center column) contrast modeled biomass with observed
ground truth. Red circles label outliers. All plots exhibit strong correlation between predicted
and observed biomass. Results in \Cref{plotJucker} (h) is best aligned with the diagonal $y=y(x)=x$
suggesting the biomass-height-diameter model as preferred. Unfortunately, in many remote sensing
scenarios estimating tree diameter is out of reach. The data in \Cref{plotJucker} (e) document the
LR2 model tend to overestimate the biomass when observed biomass is around $40$ kg, while
underestimating above about $200$ kg. In terms of height-only models, random forest and
Gaussian process regressor, \Cref{plotJucker} (k) and (n), is ruled by comparable performance
with lower bias for the full range of input data when referenced to their linear counterparts. 

The right column in \Cref{plotJucker} evaluates the density distributions of residuals $B_i - \hat{B}_i$.
Dashed lines are the $20$th (left) and $90$th (right) percentiles of the errors. Obviously, LR
and LR2 models are significantly biased with positive errors dominating. Besides, RF and GPR
distribute errors alike.

\subsection{Collected Data}
\label{sec:DataCuration}
\begin{figure*}[!t]
\centering
\subfloat[]{\includegraphics[height=1.5in]{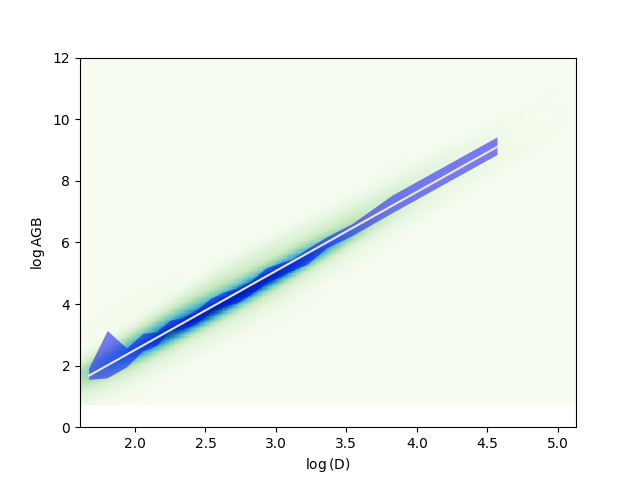}}
\subfloat[]{\includegraphics[height=1.5in]{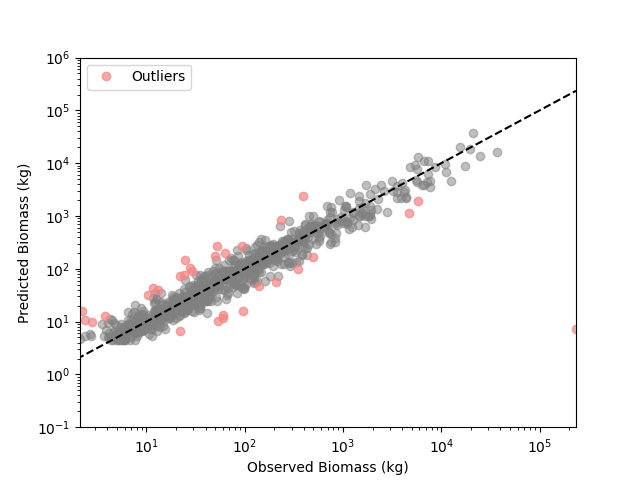}}
\subfloat[]{\includegraphics[height=1.5in]{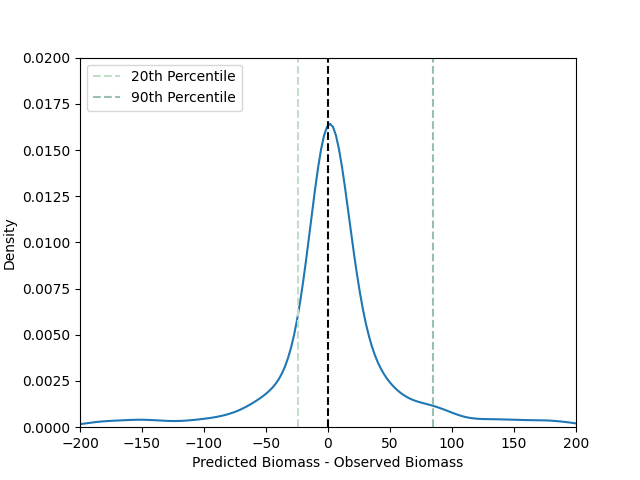}}
\hfil
\subfloat[]{\includegraphics[height=1.5in]{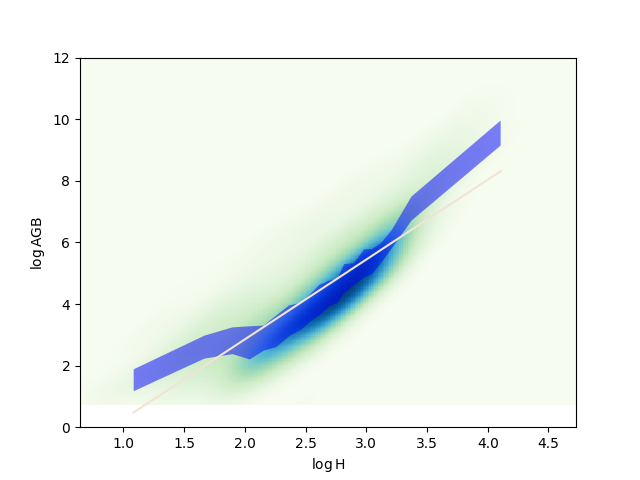}}
\subfloat[]{\includegraphics[height=1.5in]{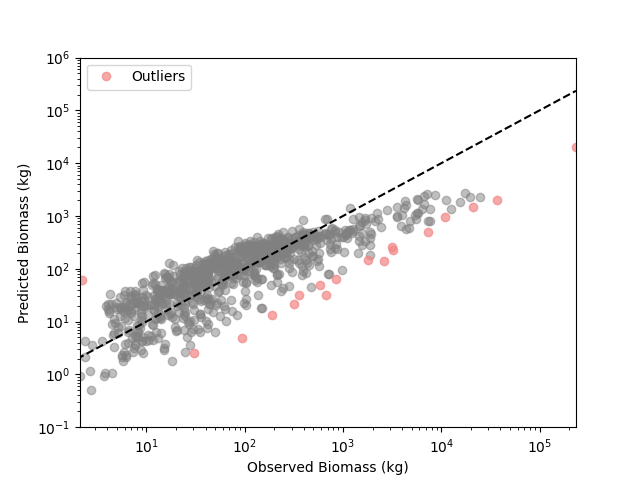}}
\subfloat[]{\includegraphics[height=1.5in]{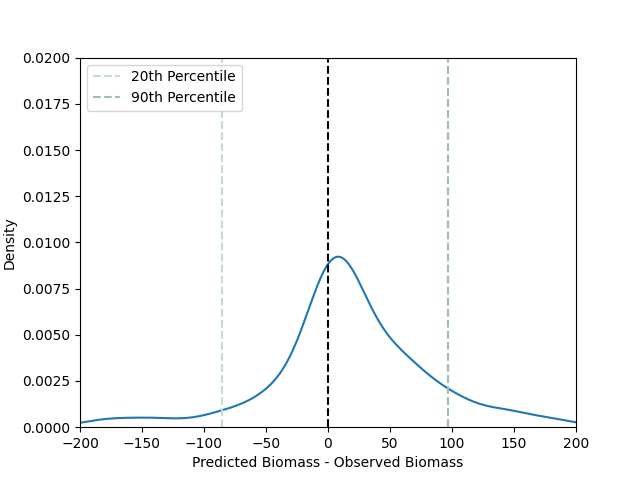}}
\hfil
\subfloat[]{\includegraphics[height=1.5in]{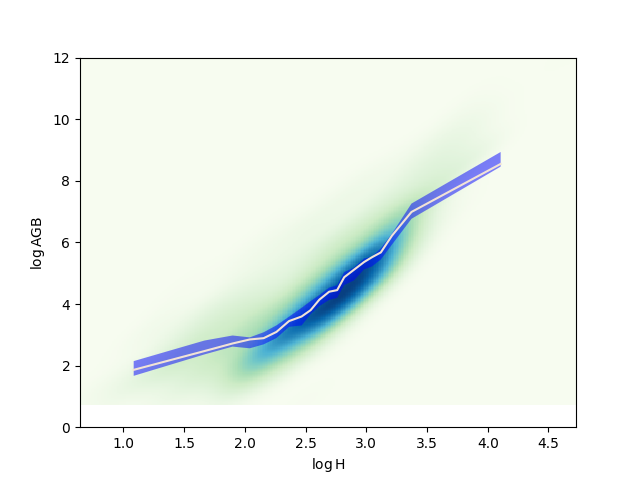}}
\subfloat[]{\includegraphics[height=1.5in]{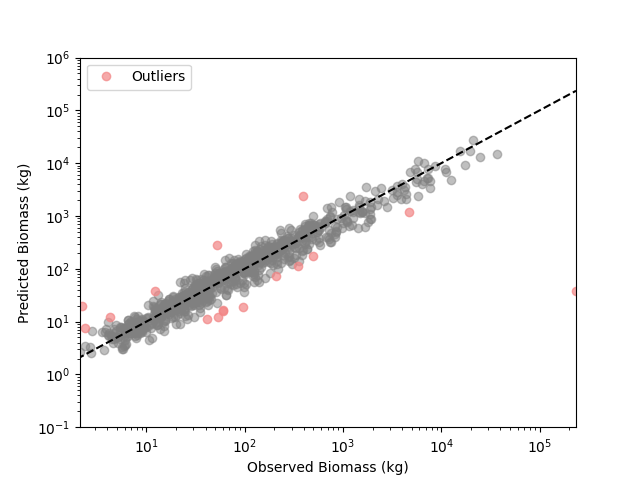}}
\subfloat[]{\includegraphics[height=1.5in]{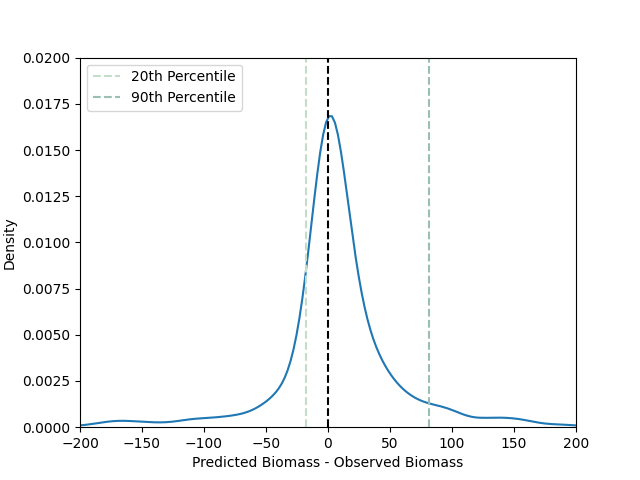}}
\hfil
\subfloat[]{\includegraphics[height=1.5in]{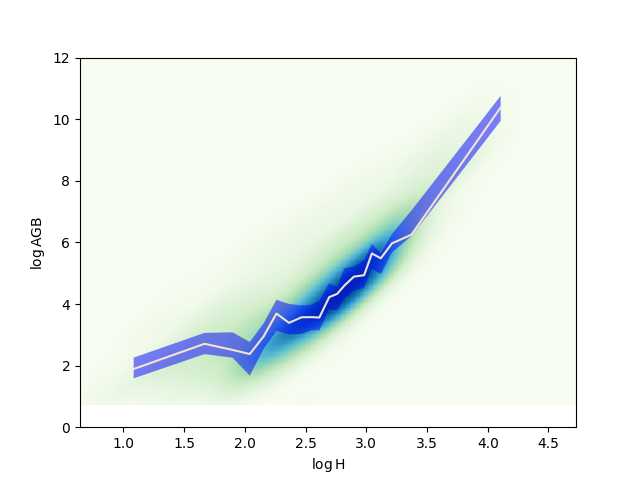}}
\subfloat[]{\includegraphics[height=1.5in]{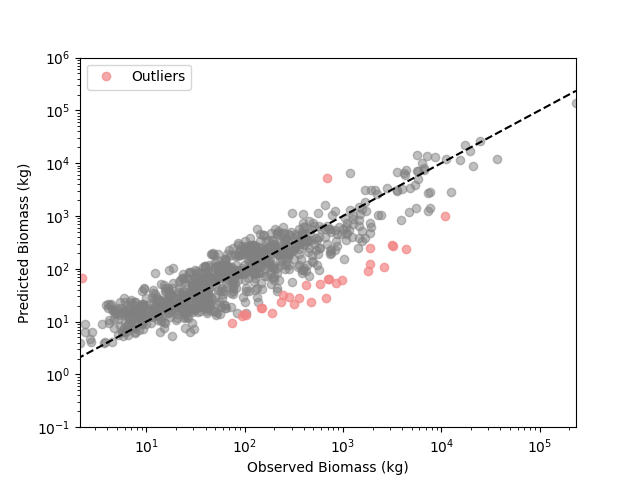}}
\subfloat[]{\includegraphics[height=1.5in]{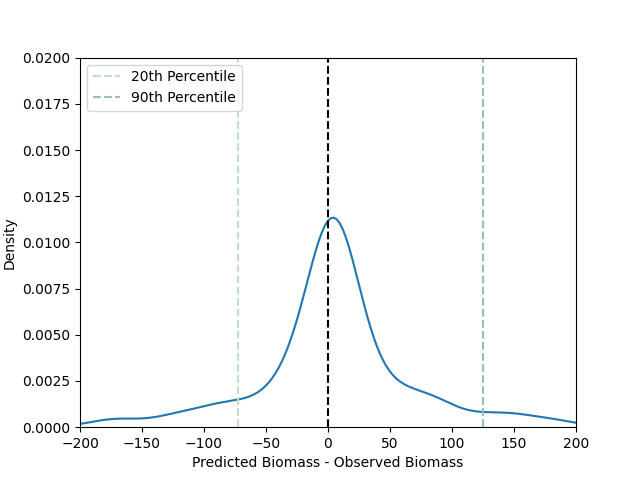}}
\hfil
\subfloat[]{\includegraphics[height=1.5in]{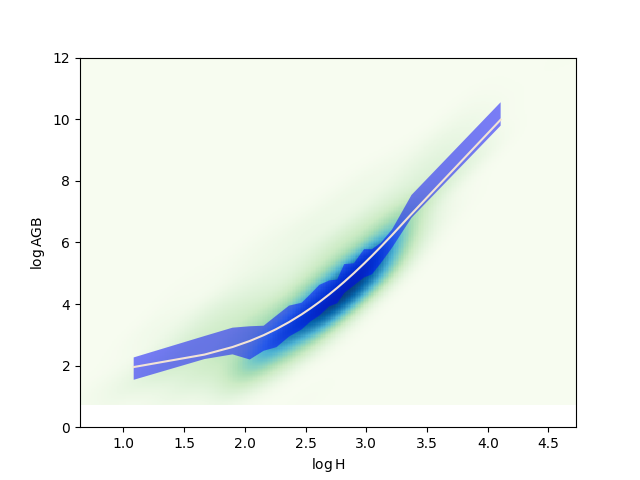}}
\subfloat[]{\includegraphics[height=1.5in]{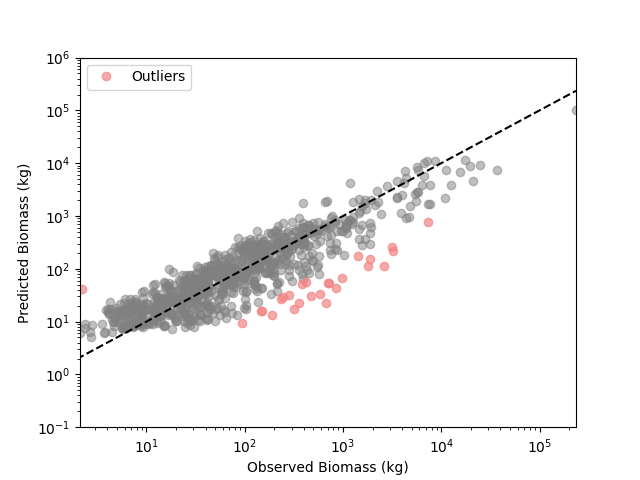}}
\subfloat[]{\includegraphics[height=1.5in]{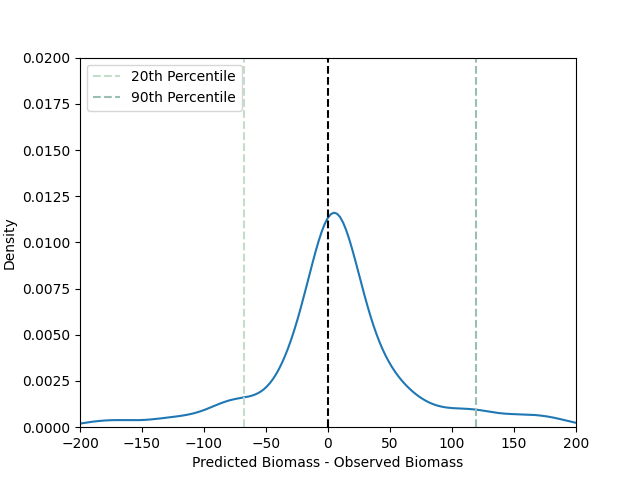}}
\caption{Plots of model fits including corresponding prediction errors (left column),
scatters of predicted and observed biomass (middle column), and the distributions of errors (right column)
based on curated data, cf.\ \Cref{datasets}. Each panel corresponds to one of the five models,
i.e.\  LR: (a)-(c), LR2: (d)-(f), LR3: (g)-(i), RF: (j)-(l), GPR: (m)-(o).}
\label{plotCollected}
\end{figure*}

The five candidate models are then trained and tested using the collected data in \Cref{datasets}. 
Since the dataset does not record crown diameter, as an alternative to LR
in Equation \ref{BHCD}, we exploit the biomass-diameter model of \Cref{B-D}.
\Cref{R2_2} and \Cref{plotCollected} present corresponding results. The plots in \Cref{plotCollected}
is arranged in line with \Cref{plotJucker}.

The LR and LR3 models yield significantly less bias---0.14 and 0.11, respectively---when compared
with the other models exceeding 0.34. $R^2$ scores for LR and LR3 read 0.73 and 0.78, respectively.
Our findings suggest that
\begin{itemize}
    \item tree diameter is relevant in biomass estimation;
    \item tree height information improves model accuracy.
\end{itemize}
It seems a linear regressor is sufficient to render the biomass--tree diameter relationship.
The dominant root mean square errors (8.2 Mg) stem from outliers (red circles) in \Cref{plotCollected} (h). 

A linear biomass-height model results in most poor performance, with $R^2$ as low as 0.25, and a
bias of 0.50. The plot in \Cref{plotCollected} (e) illustrates a significant underestimation of model
predictions vs.\ ground truth when the observed biomass exceeds 1 Mg. We conclude the log-log linear
model misses to represent the above-ground biomass--tree height relationship. In fact, the nonlinear
models outperform the linear model in terms of all three indicators. Moreover, the residual errors
in \Cref{plotCollected} (l)(o) better centers on zero compared with the results of \Cref{plotCollected}
(f); an indication of the nonlinear models more closely agree with the test data. Compared with the
random forest model the Gaussian process regressor is less biased. However, it ships with larger
$RMSE$ of 5.0 Mg and lower $R^2$ score equal to 0.66. The exceptionally high $R^2$ score roots in
top generalization ability for $B>2$ Mg. In \Cref{lidar-section} below we demonstrate that the GPR model outperforms RF. 

\begin{table}[]
\footnotesize
\renewcommand{\arraystretch}{1.35}
\caption{Summary of R-squared, RMSE, and Bias for five regression models estimating biomass from the dataset curated.}
\label{R2_2}
\centering
\setlength{\tabcolsep}{16pt}
\begin{tabular}{cccc}
\toprule
             & \textbf{R2} & \textbf{RMSE (kg)} & \textbf{Bias} \\ \hline \\[-1.2em]
\textbf{LR}  & 0.725898    & 8223.652       & 0.141245      \\
\textbf{LR2} & 0.245381    & 7722.323      & 0.504575      \\
\textbf{LR3} & 0.780377    & 8204.319      & 0.109171      \\
\textbf{RF}  & 0.812044    & 3679.73       & 0.366505      \\
\textbf{GPR} & 0.656328    & 4950.192      & 0.34472      \\
\bottomrule     
\end{tabular}
\end{table}

\subsection{Uncertainty Evaluation}
\begin{figure}[t]
\centering
\includegraphics[width=3.5in]{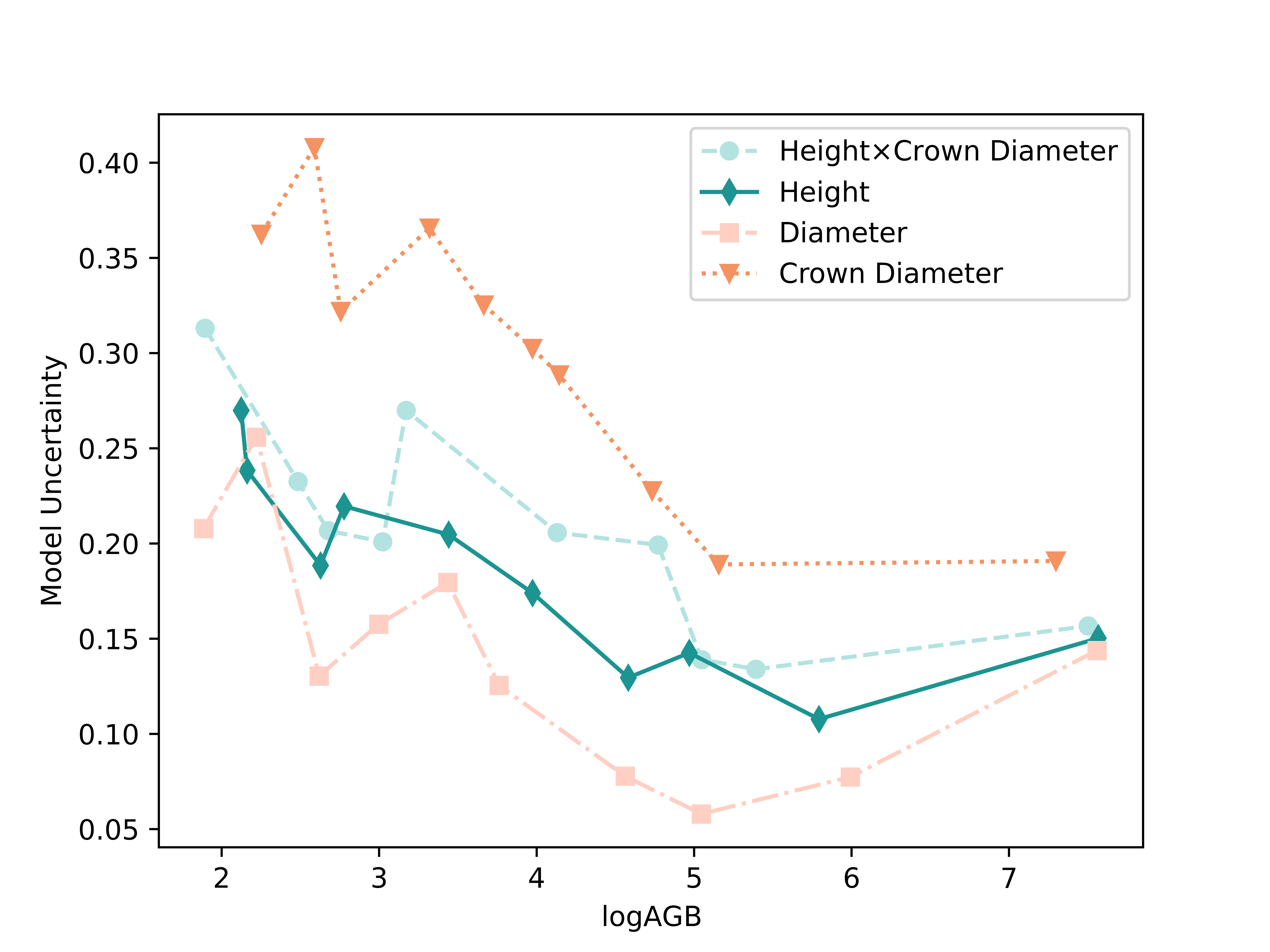}
\caption{Study of model uncertainties when working with tree height $H$, tree diameter $D$, crown diameter $CD$,
and the product $H\times CD$ as input parameter of the allometric equation.
The overall model uncertainties read: 18.25\%, 14.13\%, 29.81\%,
and 20.57\% respectively.}
\label{modelUncertainty}
\end{figure}

We consult the Jucker data \cite{Jucker} to quantify model uncertainties. \Cref{modelUncertainty}
contrasts the model uncertainties of the four models: biomass-height, biomass-diameter,
biomass-crown diameter, and biomass-height-crown diameter, respectively. It suggests that diameter
is closely related to biomass. The biomass-diameter model exhibits the lowest model uncertainty
of about 14\%. The biomass-height model reaches medium performance at overall model uncertainty
of 18.25\%. The overall model uncertainty of the single-parameter biomass-crown diameter model
is 30\%. 
As a result, the biomass-height-crown diameter relationship---cf.\ the LR model in
\Cref{BHCD}---is plagued by major model uncertainty of about 20.6\%.
Here we contrast single parameter models, only. Multiple-parameter models, such as the
biomass-height-diameter model of \Cref{BHD}, reduce uncertainty.

\begin{figure}[t]
\centering
\includegraphics[width=3.5in]{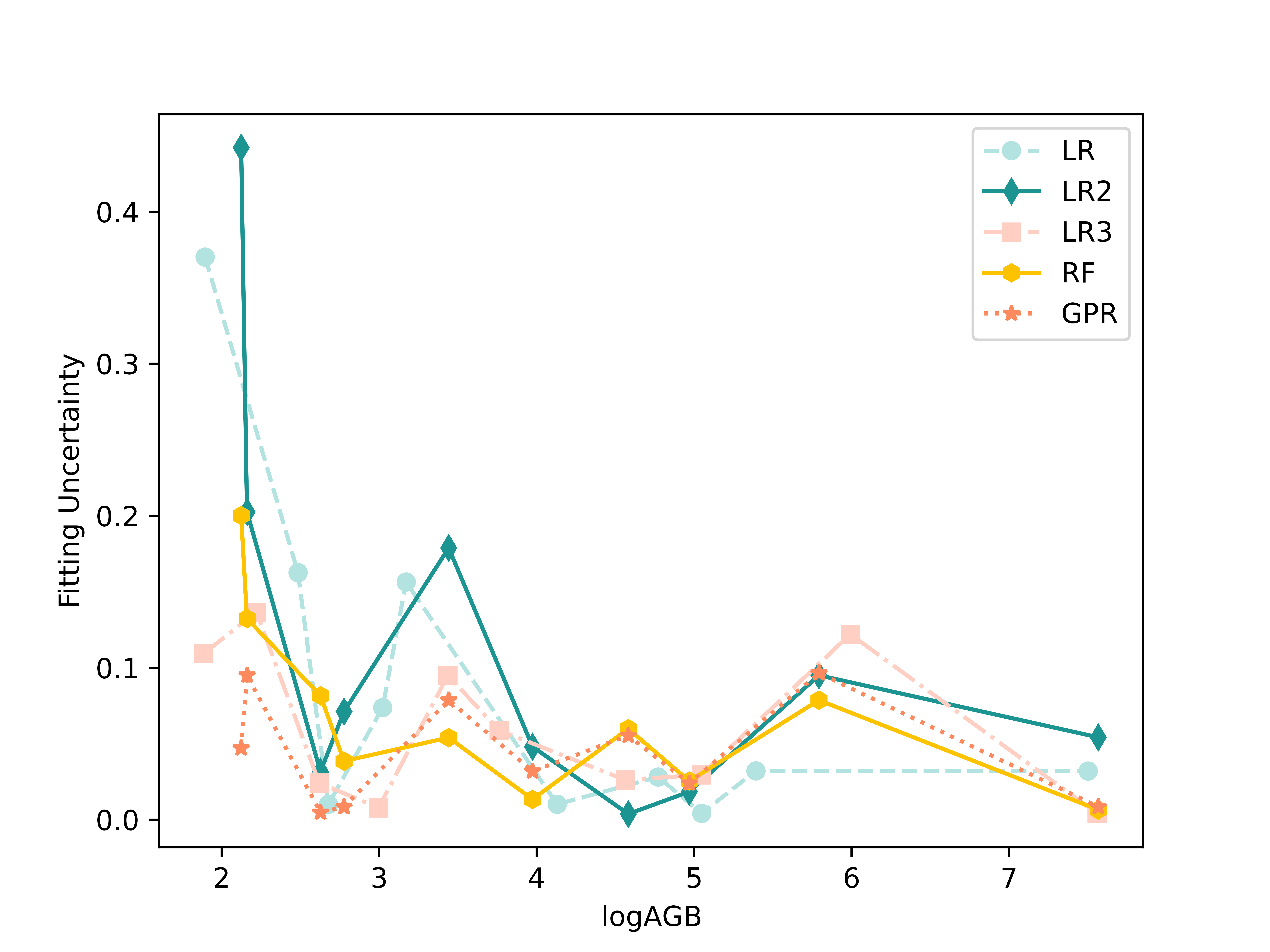}
\caption{Biomass-dependent fitting uncertainties for the five candidate models trained on the
Jucker data. The overall fitting uncertainties of the five candidate models reduce to 8.80\%,
11.45\%, 6.13\%, 6.90\%, and 4.50\% respectively.}
\label{fitUncertainty}
\end{figure}

\Cref{fitUncertainty} aggregates fitting uncertainties of our five candidate biomass models learned
from the Jucker data. In general, fitting uncertainties stay below model uncertainties. The overall
fitting uncertainties of LR, LR2, LR3, random forest, and the proposed Gaussian process regressor
read 8.80\%, 11.45\%, 6.13\%, 6.90\%, and 4.50\% respectively. We conclude estimation errors is dominated
by model uncertainty: All five models exhibit higher fitting uncertainty when the observed biomass is
less than 2.5 $\log$ kg) with the GPR model (marked by star) performing best. Because of the non-linear
biomass-height relationship, the LR2 model (marked by diamond) scores highest with respect to fitting uncertainty,
LR indicates medium performance, while LR3 and RF unveil performance scores on equal level.
The GPR model demonstrates the lowest overall fitting uncertainty. Moreover, it constantly performs in
all the groups suggesting the Gaussian process regressor over the other models in terms of low
fitting uncertainty.

%---------------------------------LiDAR------------------------------------
\section{Validation by LiDAR Data}
\label{lidar-section}
Finally, we study the uncertainty of trained models on stand level. We utilize a dataset
that get assembled from forests in Baden-Württemberg, Germany in the years 2019 and 2020
\cite{weiser2021tuaa}. It embraces 12 separate plots, each covering a spatial area of about
one hectare. For each plot, point clouds of individual trees get segmented from
terrestrial, UAV-borne and airborne LiDAR devices. Field inventory measurements are available
for a fraction of trees, too. We exclude from the validation process three out of the 12
plots with less than 20 trees available. 

\begin{figure}[t]
\centering
\subfloat[]{\includegraphics[width=3.5in]{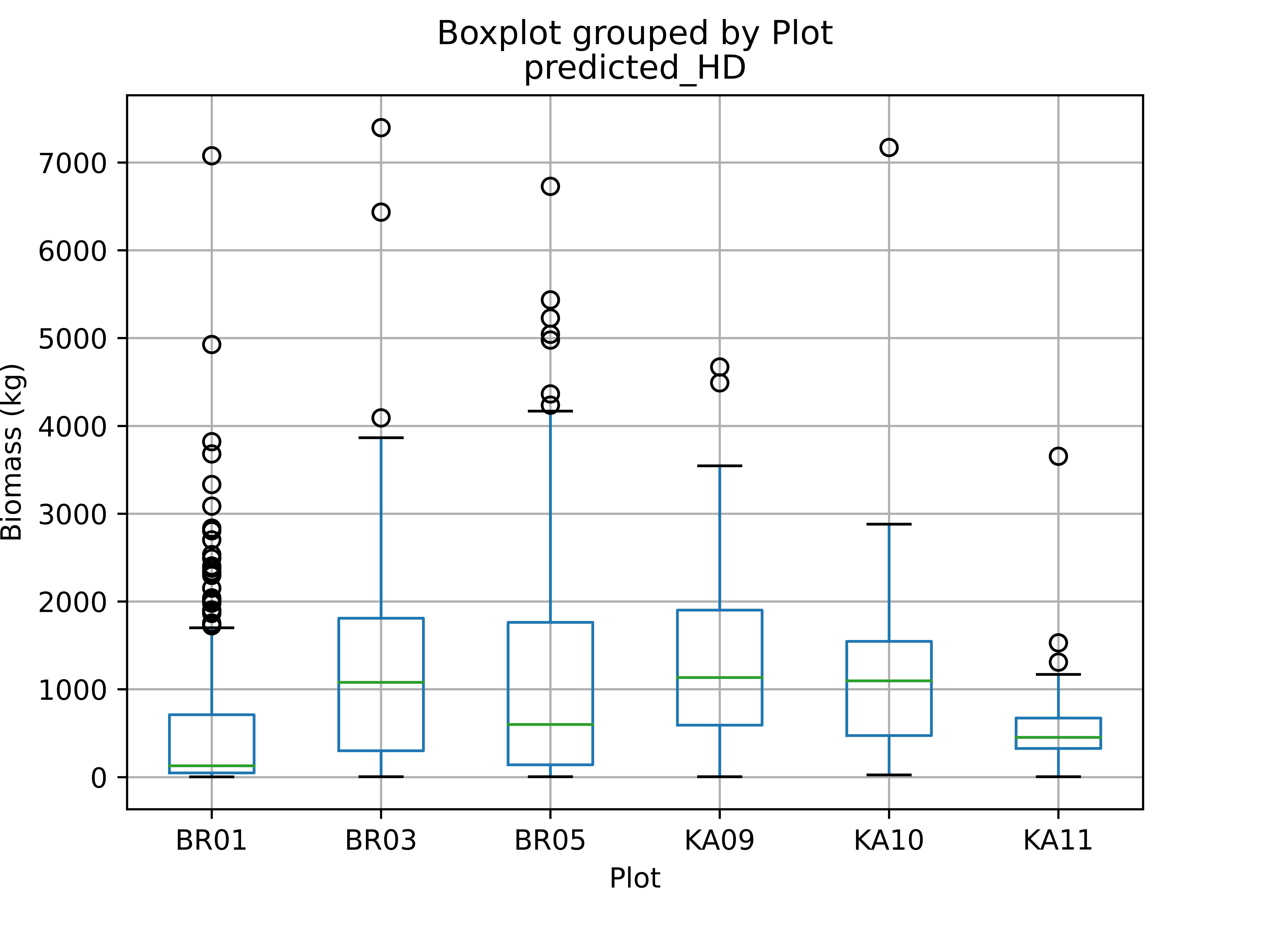}}
\hfil
\subfloat[]{\includegraphics[width=3.5in]{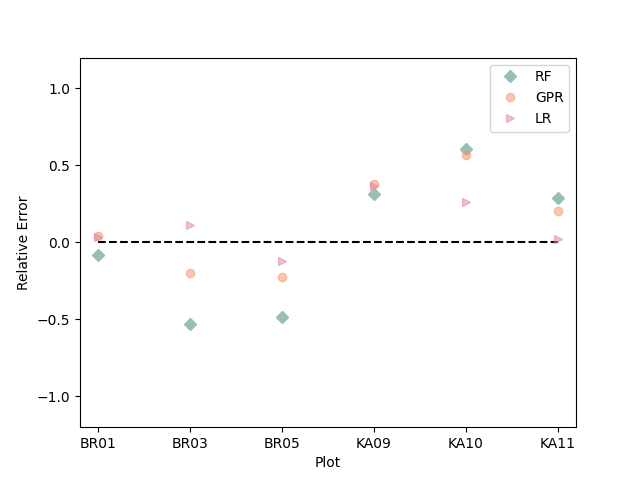}}
\hfil
\subfloat[]{\includegraphics[width=3.5in]{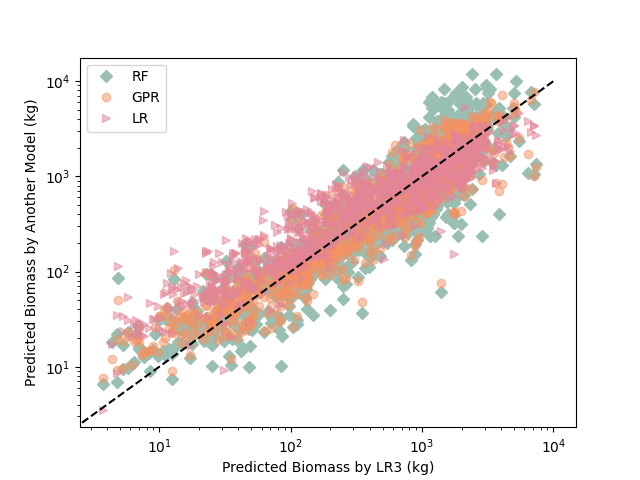}%
}
\caption{(a): Box plot of predicted biomass by LR3 model on individual level grouped by plot;
(b): the relative error of the three models in each plot where the LR3 model is assumed ground
truth; (c): scatter plot of model predicted biomass versus LR3 model reference.}
\label{val_lidar}
\end{figure}

For each tree, its height, diameter, and crown diameter are derived either by field
measurements and LiDAR point cloud data, and discard from the analysis trees without field measurements. For a single tree, there may exist multiple LiDAR
measurements and the number of measurements fluctuates from plot to plot. Therefore, we average all measurements. Note that these measurements are incomplete, for example, tree heights were not inventoried (or measured by LiDAR data). In those cases, LiDAR-measured (or inventoried) variables were used instead. In \Cref{val_lidar} (a), we present box plots of the tree biomass
grouped by stands where it is indicated the following biomass statistics
from bottom to top: minimum, first quartile, median, third quartile, and maximum.
Outliers get represented by black, empty circles. The tree biomass value dominantly varies
from zero to two Mg. Its distributions exhibit distinct characteristics for each plot.
We note that the fluctuating number of outliers has the potential to impact model uncertainty.

The computation serves as a basis to compare three candidate models: LR from \Cref{BHCD}
trained on Jucker data, and models RF \& GPR trained on the data curated in \Cref{sec:DataCuration}.
In order to derive stand-level biomass estimates, we sum up the values of individual trees.
This way, the
over- and underestimation of biomass, in large parts, cancel. 
For each plot \Cref{val_lidar} (b) presents the resulting RE values of the three models RF, GPR, and LR.
RF most poorly performs for all plots except KA09. However, LR underestimates the biomass
in seven out of nine plots, thus there is a significant bias for errors to accumulate.
In \Cref{val_indicators} we compute the relative error of the three models given all reference
data. We notice the LR model rendering more
biased compared with the GPR. The scatter plot in \Cref{val_lidar} (c) supports that LR tends
to over- and underestimate tree biomass when assuming values less and larger than 500 kg,
respectively. GPR-predicted biomass estimates well correlate with the predicted biomass of
the LR3 model reference.

As listed in \Cref{val_indicators}, the relative RMSE of models
LR, RF, and GPR assume values 16.93\%, 46.08\%, and 24.46\%, respectively. Although GPR is less accurate compared with the LR model, the relative RMSE of GPR is acceptable when
referenced to the state-of-the-art biomass estimation errors on a national and global scale,
cf.\ \cite{wall-to-wall} quoting \%RMSE values in 37\% to 67\%.

\begin{table}[]
\renewcommand{\arraystretch}{1.33}
\caption{Comparison of model performance in terms of relative error and relative RMSE for
candidate models: LR, RF, and GPR}
\label{val_indicators}
\centering
\setlength{\tabcolsep}{16pt}
\begin{tabular}{cccc}
\toprule
                & \textbf{LR} & \textbf{RF} & \textbf{GPR} \\ \hline
\textbf{RE}     & 0.0677                  & -0.1712             & 0.0021                            \\
\textbf{\%RMSE} & 0.1693         & 0.4608                & 0.2446                           \\  \bottomrule    
\end{tabular}
\end{table}

%------------------------------------------conclusion------------------------------------------
\section{Conclusion}
\label{conclusions}
We proposed a Gaussian process regressor model to estimate biomass on individual tree level
taking tree height as input, only. It enables rapid regional-to-national above-ground biomass evaluation from high-resolution LiDAR data. As a single input parameter model, a series of
existing allometry databases contribute to model training. We benchmarked GPR against four
established biomass models training on Jucker data and a dataset curated by this work.
Results confirm GPR performs best when compared with two biomass-height models, and it achieves
reasonable results in reference to a biomass-height-crown diameter model. GPR generates a low
fitting uncertainty of 4.50\%. Stand-level uncertainty analysis of GPR yielded an averaged relative
root mean square error of 24.46\%. Moreover, GPR renders less biased at a mean relative error of 0.0021.
Future work may explore a stratified approach where biome-specific models \cite{Jucker} have
the potential to decrease model uncertainty. 

%\section*{Acknowledgment}

\ifCLASSOPTIONcaptionsoff
\newpage
\fi

\bibliographystyle{ieeetr}
%\bibliography{reference}

\begin{IEEEbiography}[{\includegraphics[width=1in, height=1.25in,clip,keepaspectratio]{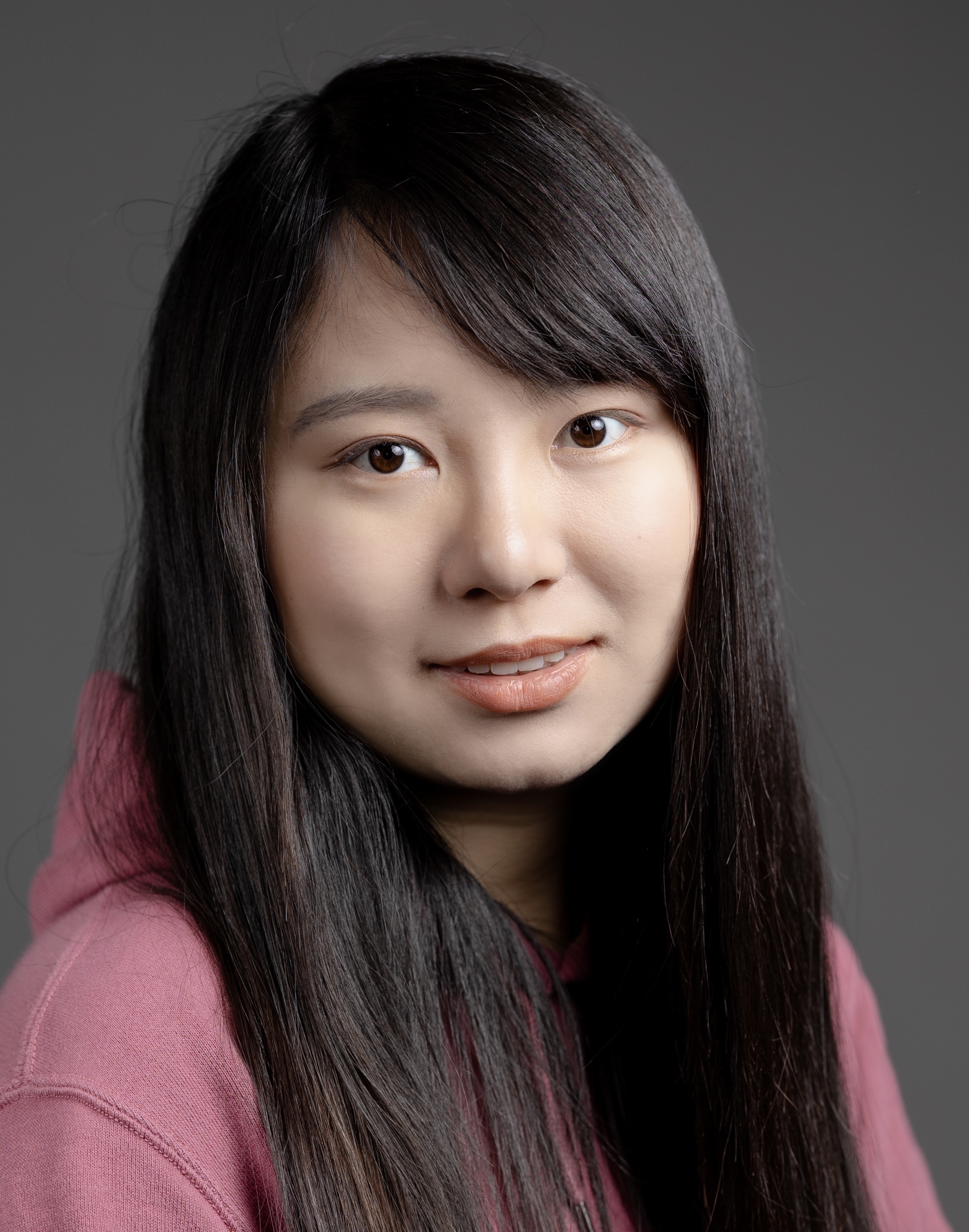}}]{Qian Song}
(S'16-M'20) received the B.E. degree (Hons.) from the School of Information Science and Technology, East China Normal University, Shanghai, China, in 2015, and the Ph.D. degree (Hons.) from Fudan University, Shanghai, China, in 2020. She was a post-doctoral fellow with the Remote Sensing Technology Institute (IMF), German Aerospace Center (DLR), Wessling, Germany from 2020 to 2022. Since 2023, she is a post-doctoral fellow with the Chair of Data Science in Earth Observation at the Technical University of Munich (TUM), Munich, Germany.

She has been awarded as the URSI (International Union of Radio Science) Young Scientist Award in 2020. Her research interests include advanced deep learning technologies and their applications in synthetic aperture radar image interpretation, forest monitoring and biomass estimation.
\end{IEEEbiography}

\begin{IEEEbiography}[{\includegraphics[width=1in,height=1.25in,clip,keepaspectratio]{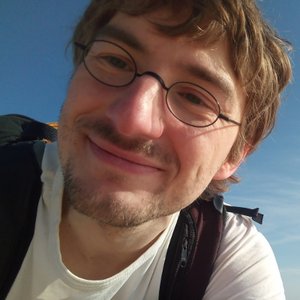}}]{
Conrad M Albrecht} (M'17) received an undergraduate degree in physics from Technical
University Dresden, Germany, in 2007 and a Ph.D. degree in physics with an extra
certification in computer science from Heidelberg University, Germany, in 2014.
Spanning the fields of physics, mathematics, and computer science, among others,
he was a visiting scientist with CERN, Switzerland, in 2010, and with the Dresden
Max Planck Institute for the Physics of Complex Systems, Germany, in 2007.

In 2015 Conrad became a research scientist with the IBM T.J. Watson Research Center,
Yorktown Heights, NY, USA. Currently, since April 2021, he leads a HelmholtzAI-funded
team for "Large-Scale Data Mining in Earth Observation" (DM4EO) at the German Aerospace Center,
Oberpfaffenhofen, Germany in close collaboration with Technical University Munich.
Starting 2023, Conrad's team contributes to Horizon Europe project EvoLand, \url{https://www.evo-land.eu}.

Conrad's research agenda interconnects physical models and numerical analysis, employing
Big Data technologies and machine learning. As part of the "Data Intensive Physical
Analytics" team in IBM Research, he significantly contributed to industry-level solutions
processing geospatial information with focus on machine-learning driven remote sensing
applications. He co-organized workshops at the IEEE BigData conference, the EGU General Assembly,
and the AAAS annual meeting. Home to the US and the EU, Conrad's scientific agenda aims to strengthen
transatlantic collaboration of corporate research and academia. His DM4EO team collaborates
with institutions such as Juelich Super-Computing Center, GFZ German Research Centre for Geosciences,
INRIA Grenoble, Princeton University, IBM Research, and Yale University.\end{IEEEbiography}

\begin{IEEEbiography}[{\includegraphics[width=1in, height=1.25in,clip,keepaspectratio]{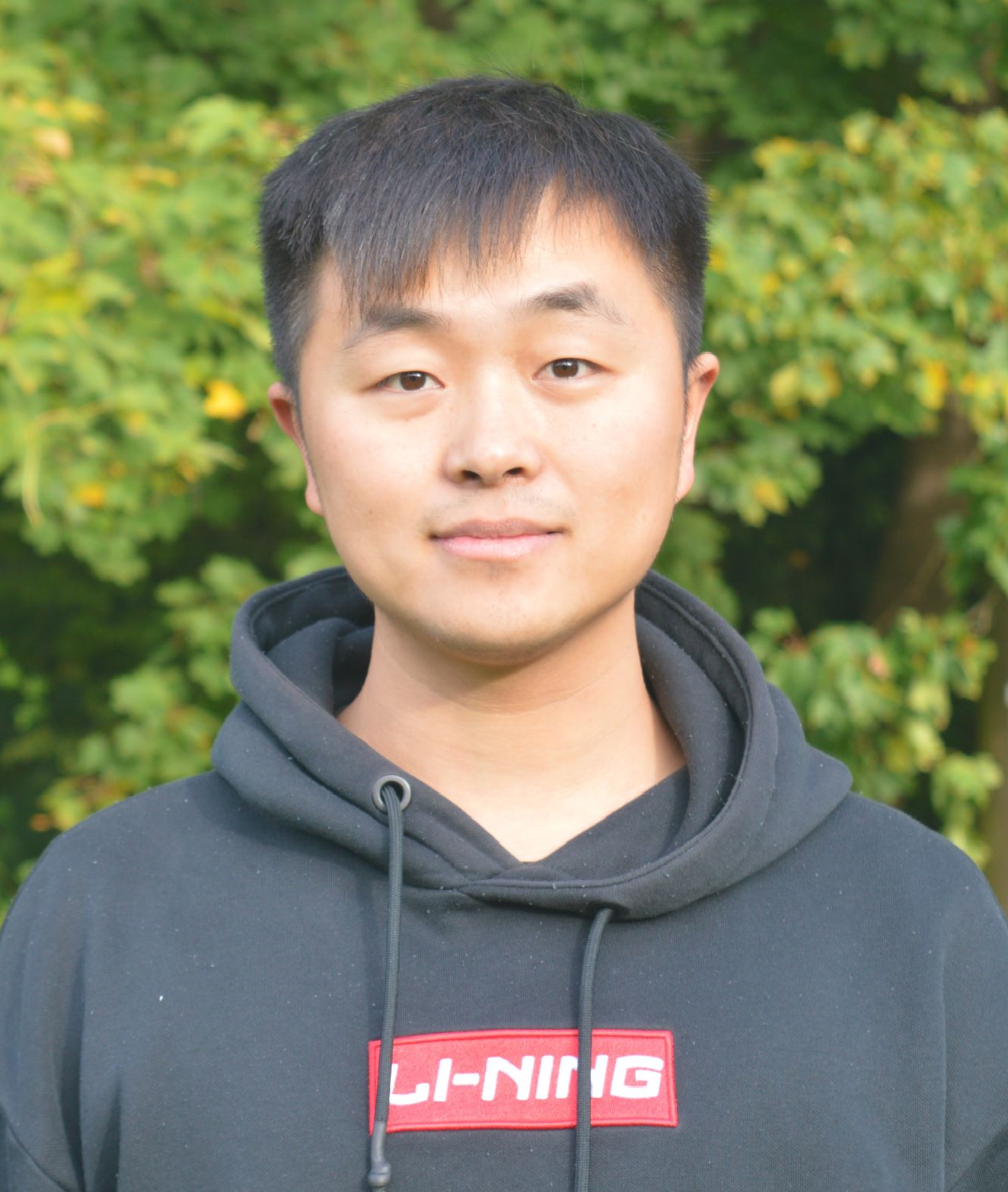}}]{Zhitong Xiong} received the Ph.D. degree in computer science and technology from Northwestern Polytechnical University, Xi’an, China, in 2021. He is currently a senior scientist and leads the ML4Earth working group with the Data Science in Earth Observation, Technical University of Munich (TUM), Germany. His research interests include computer vision, machine learning, Earth observation, and Earth system modeling.
\end{IEEEbiography}

\begin{IEEEbiography}[{\includegraphics[width=1in,height=1.25in,clip,keepaspectratio]{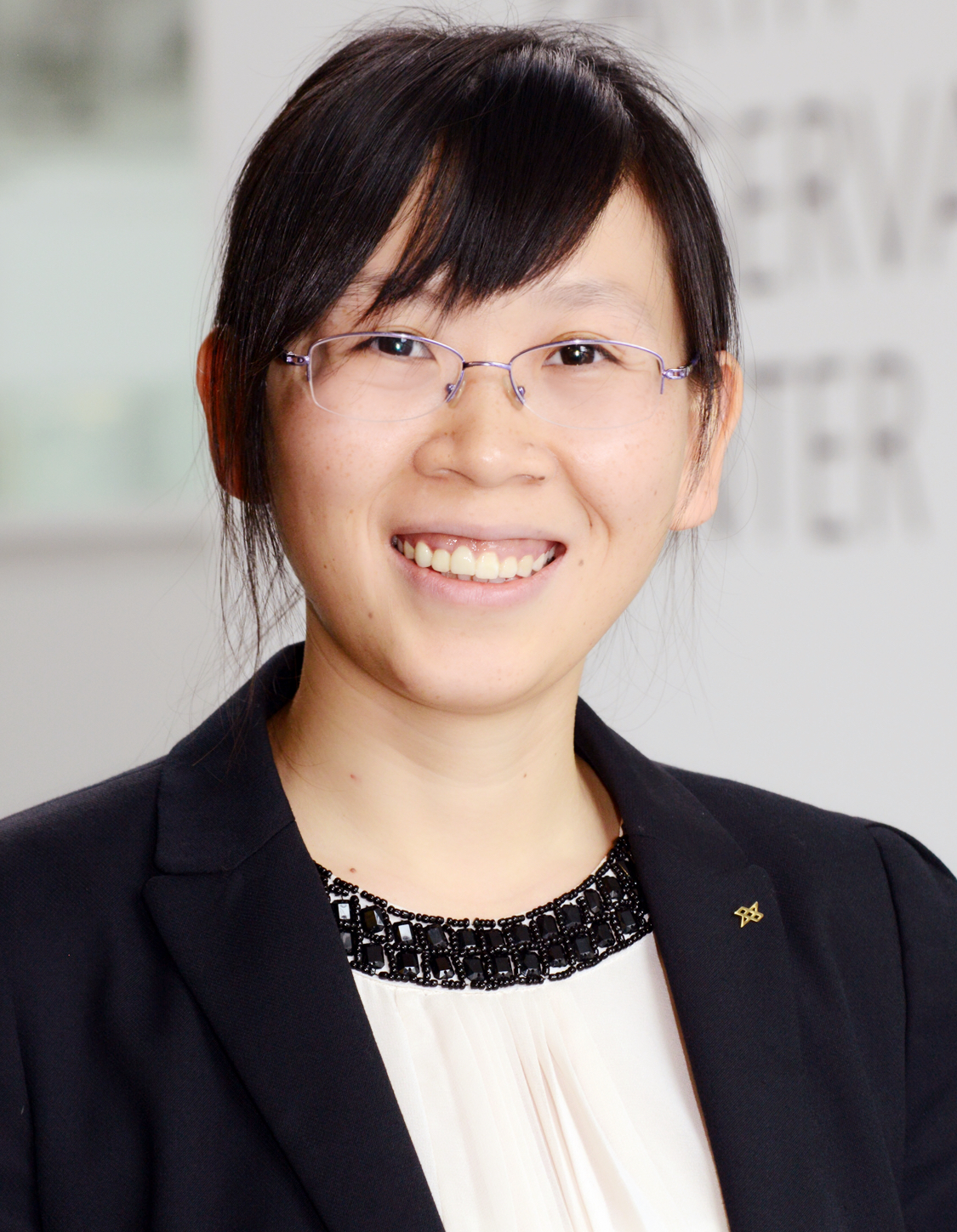}}]{Xiao Xiang Zhu}(S'10--M'12--SM'14--F'21) received the Master (M.Sc.) degree, her doctor of engineering (Dr.-Ing.) degree and her “Habilitation” in the field of signal processing from Technical University of Munich (TUM), Munich, Germany, in 2008, 2011 and 2013, respectively.
\par
She is the Chair Professor for Data Science in Earth Observation at TUM and was the founding Head of the Department ``EO Data Science'' at the Remote Sensing Technology Institute, German Aerospace Center (DLR). Since 2019, Zhu is a co-coordinator of the Munich Data Science Research School (www.mu-ds.de). Since 2019 She also heads the Helmholtz Artificial Intelligence -- Research Field ``Aeronautics, Space and Transport". Since May 2020, she is the PI and director of the international future AI lab "AI4EO -- Artificial Intelligence for Earth Observation: Reasoning, Uncertainties, Ethics and Beyond", Munich, Germany. Since October 2020, she also serves as a Director of the Munich Data Science Institute (MDSI), TUM. Prof. Zhu was a guest scientist or visiting professor at the Italian National Research Council (CNR-IREA), Naples, Italy, Fudan University, Shanghai, China, the University  of Tokyo, Tokyo, Japan and University of California, Los Angeles, United States in 2009, 2014, 2015 and 2016, respectively. She is currently a visiting AI professor at ESA's Phi-lab. Her main research interests are remote sensing and Earth observation, signal processing, machine learning and data science, with their applications in tackling societal grand challenges, e.g. Global Urbanization, UN’s SDGs and Climate Change.

Dr. Zhu is a member of young academy (Junge Akademie/Junges Kolleg) at the Berlin-Brandenburg Academy of Sciences and Humanities and the German National  Academy of Sciences Leopoldina and the Bavarian Academy of Sciences and Humanities. She serves in the scientific advisory board in several research organizations, among others the German Research Center for Geosciences (GFZ) and Potsdam Institute for Climate Impact Research (PIK). She is an associate Editor of IEEE Transactions on Geoscience and Remote Sensing and serves as the area editor responsible for special issues of IEEE Signal Processing Magazine. She is a Fellow of IEEE.
\end{IEEEbiography}

\end{document}